\documentclass[pdflatex,sn-mathphys]{sn-jnl}
\usepackage{bm}
\usepackage[utf8]{inputenc}
\usepackage[T1]{fontenc}
\jyear{2024}%
\theoremstyle{thmstyleone}%
\theoremstyle{thmstyletwo}%
\theoremstyle{thmstylethree}%
\DeclareMathOperator*{\argmin}{arg\,min}
\begin{document}
\title[A Systematic Review of Local Low-Rank Approximation in Medical Imaging]{A Systematic Review of Low-Rank and Local Low-Rank Matrix Approximation in Big Data Medical Imaging}
\author*[1,2]{Sisipho Hamlomo}
\author*[1]{Marcellin Atemkeng}
\author[3]{Yusuf Brima}
\author[4,5,6]{Chuneeta Nunhokee}
\author[2]{Jeremy Baxter}
\affil[1]{\small Department of Mathematics, Rhodes University, PO Box 94, Makhanda, 6140, South Africa}
\affil[2]{\small Department of Statistics, Rhodes University, PO Box 94, Makhanda, 6140, South Africa}
\affil[3]{\small Computer Vision, Institute of Cognitive Science, Osnabr\"uck University, Wachsbleiche 27, Osnabr\"uck, D-49090, Germany}
\affil[4]{\small International Centre for Radio Astronomy Research (ICRAR), Curtin University, Bentley, WA, Australia}
\affil[5]{\small ARC Centre of Excellence for All Sky Astrophysics in 3 Dimensions (ASTRO 3D), Bentley, Australia}
\affil[6]{\small Curtin Institute of Radio Astronomy, GPO Box U1987, Perth, WA 6845, Australia}
\affil[*]{\small Correspondence: s.hamlomo@ru.ac.za, m.atemkeng@ru.ac.za}

\abstract{
The large volume and complexity of medical imaging datasets pose significant challenges for storage, transmission, and processing. To address these issues, low-rank matrix approximation (LORMA) and its derivative, local LORMA (LLORMA), have shown promising potential. This paper presents a comprehensive literature review of the application of LORMA and LLORMA across various imaging modalities and examines the challenges and limitations of existing methods. Notably, since 2015, there has been a significant shift toward a preference for LLORMA in the medical imaging field, demonstrating its effectiveness in capturing complex structures in medical data compared to LORMA. Given the limitations of shallow similarity methods in LLORMA, we propose incorporating advanced semantic image segmentation to improve the accuracy of similarity measurement. We further explain how this approach can be utilized to identify similar patches and assess its feasibility in medical imaging applications.

We observe that LORMA and LLORMA have primarily been applied to unstructured medical data, and we suggest extending their use to other types of medical data, including structured and semi-structured formats. This paper also explores how LORMA and LLORMA can be adapted for regular data with missing entries, considering the impact of inaccuracies in predicting these missing values and their consequences. In addition, we examine the effect of patch size and suggest using random search (RS) to identify the optimal patch size. To further enhance feasibility, we propose a hybrid approach combining Bayesian optimization and RS, which could improve the application of LORMA and LLORMA in medical imaging.
}





\maketitle
\section{Introduction}
Medical images obtained using technologies such as PET, CT scans, MRI and ultrasound are crucial for diagnosing and monitoring health conditions \cite{hussain2022modern, umar2019review,ganguly2010medical}. These images are central to treatment planning, guiding surgical procedures, and facilitating minimally invasive procedures \cite{shah2014recent,lieberman2006bone}. They also play a crucial role in monitoring disease progression, assessing treatment effectiveness and making timely adjustments to the treatment plan 
\cite{umar2019review, islam2023introduction, tavazzi2020mri, delorme2009imaging}. Rapid and accurate diagnosis through medical imaging is crucial in emergency medicine for identifying critical conditions and initiating prompt intervention \cite{chin2013pilot, tayal2003prospective}. In addition to clinical applications, medical images play a vital role in research, education, and the emerging field of personalized medicine, where treatments tailored to individual anatomy and pathology are becoming increasingly important \cite{aerts2016potential, maceachern2021machine,pinker2018precision}. However, managing medical data is challenging due to issues such as noise, high-dimensionality, corrupted or incomplete data, and large volumes of data, which can increase memory requirements and computational time \cite{li2018large,panayides2020ai,yu2021convolutional, bolon2016feature, lee2017medical}. Various techniques have been proposed in the literature to address these challenges. These include smoothing algorithms (i.e., Weiner filter, Gaussian filter, average filter, median filter, etc.) used to process and analyze medical datasets to reduce noise \cite{shinde2012noise,na2018guided,kumar2017noise, ilango2011new,bhonsle2012medical,gravel2004method,naimi2015medical}. Methods like sparse coding \cite{devadoss2019near,sudha2011two,liu2019fast,anandan2016medical}, discrete cosine transform (DCT) and wavelet transform are commonly employed for image compression \cite{jones1995comparative,wu2001medical,kaur2015image,singh2015multiple}. Additionally, techniques such as t-distributed stochastic neighbor embedding (t-SNE) and local linear embedding (LLE) are used to reduce the dimensions of these datasets \cite{devassy2020dimensionality,bibal2020explaining,faust2018visualizing,jamieson2010exploring}. However, a common limitation of these smoothing algorithms is their inability to effectively handle images corrupted by complex or non-Gaussian noise patterns. These filters often struggle to preserve fine details and edges in noisy images, which can result in over-smoothing or incomplete noise removal \cite{jin2003adaptive,milanfar2013symmetrizing,doymaz2001wavelet,rahman2003wavelet}. Furthermore, DCT and wavelet transform techniques can introduce blocking and ringing artifacts and fail to capture directional information \cite{pradhan2016comparative,toufik2012wavelet}. While sparse coding is effective, it can be computationally expensive, especially for large images \cite{xiang2011learning,zheng2010graph}. Moreover, the computational and space complexity of t-SNE grows quadratically with the number of data pairs, limiting its application to large datasets \cite{van2008visualizing,gisbrecht2015parametric,van2014accelerating}. Determining an appropriate $k-$nearest neighbors value for all input data in LLE is also challenging due to the complexity, nonlinearity, and diversity of high-dimensional input data \cite{saul2003think,ge2008hand,pan2009weighted}.

Low-rank matrix approximation (LORMA) has shown potential in addressing the challenges mentioned above \cite{zhou2014low, gu2019brief, haeffele2014structured,ebiele2020conventional}. LORMA refers to a technique that approximates a given matrix as the product of two lower-dimensional matrices, reducing the complexity of the original matrix while preserving important features. These techniques allow these large datasets to be compressed and efficiently represented, enabling faster storage, transmission, and processing \cite{lee2017deep,muller2004review,razzak2018deep}. However, these datasets may be affected by missing or corrupted entries due to factors such as equipment limitations or patient movement \cite{magrabi2012using,shellock2002magnetic,catana2015motion}. LORMA techniques, such as image inpainting, are employed to reconstruct missing or corrupted regions of the medical images \cite{wang2021medical,armanious2020ipa,li2019progressive}. 
In signal processing, particularly in biological signal analysis from physiological measurements, LORMA techniques assist with denoising and feature extraction, thereby improving diagnostic accuracy \cite{sweeney2012artifact,singh2006optimal}. In genomics and proteomics, LORMA is used for data analysis, facilitating dimensionality reduction and pattern recognition in complex biological datasets \cite{meng2016dimension, misra2019integrated}. In chemoinformatics, LORMA aids in the analysis of chemical and biological data matrices in drug discovery, helping to identify relationships between molecular structures and biological activities \cite{hao2018new,t20165}. In multi-modal data fusion -- when dealing with data from multiple sources (e.g., imaging, clinical, genomic), LORMA is utilized to integrate and fuse different modalities, providing a comprehensive view for diagnosis and treatment planning \cite{he2022fidelity,liu2015medical,zhong2016predict}.

While LORMA proves to be a valuable tool in medical imaging, it has limitations. One of the primary concerns is the potential loss of image detail and fidelity \cite{huang2014robust,lu2016reference,lu2018gradient}. As the rank of the approximation decreases, fine textures, intricate patterns, and subtle color variations may be compromised, leading to a degraded image. The choice of the appropriate rank becomes crucial, striking a balance between reducing storage requirements and preserving visual quality \cite{oseledets2009breaking, hastie2015matrix}. Another challenge lies in the computation time associated with LORMA methods. Singular value decomposition (SVD), a common technique for LORMA, has a time complexity of approximately $\mathcal{O}(mn\min\{m,\,n\})$ \cite{holmes2007fast, mamat2007statistical, brand2006fast, roughgarden2015cs168, atemkeng2023lossy, halko2011finding}, where $m$ and $n$ are the dimensions of the image and truncating it to $r$ singular values and vectors results in a complexity of $\mathcal{O}(mnr)$ \cite{brand2006fast, roughgarden2015cs168, burca2014fast, atemkeng2023lossy, halko2011finding}. This can be computationally expensive for high-dimensional images. While randomized SVD offers a faster alternative with a time complexity of $\mathcal{O}(mn\log(r))$ for rank-$r$ approximation \cite{halko2011finding}, the computation still demands careful consideration, especially for real-time applications.

The local low-rank matrix approximation, an extension of LORMA proposed in \cite{lee2013local} in 2013 overcomes the drawbacks mentioned above and significantly relaxes the low-rank assumption. Rather than assuming that a matrix is globally low-rank, this approach assumes that the matrix behaves as a low-rank matrix within the neighborhood of specific row-column combinations. This enables the capture of complicated spatial patterns in the data. By decomposing the input image into low-rank components specific to local regions, the technique captures local variations and fine-grained details that may be missed in global LORMA \cite{lee2013local, liu2019adaptive}. This makes it particularly suitable for image processing tasks, where objects and features often exhibit varying degrees of complexity and structure across different regions \cite{guo2017patch}. By exploiting the redundancy and similarity of neighboring regions, this technique reduces the computational complexity associated with traditional LORMA \cite{lee2013local, dib2020local}. The resulting efficiency can lead to faster processing times and reduced memory requirements, making it feasible to handle larger datasets and real-time applications \cite{lee2016llorma}.

However, the LLORMA technique has its drawbacks. One notable limitation is the potential for patch artifacts or inconsistencies along the boundaries of localized regions. These artifacts can occur due to abrupt transitions between adjacent regions with different low-rank structures. Careful selection of patch sizes and strategies to mitigate such artifacts are essential to maintaining the quality of the approximation. Additionally, for certain types of data where information from distant regions is critical to the overall structure, LLORMA may struggle to capture global dependencies and long-range interactions.

This review highlights works in the literature that apply LORMA and LLORMA to medical images. To the best of our knowledge, no systematic literature review specifically focuses on LORMA and/or LLORMA in the context of medical data. In light of this gap, review questions are formulated to guide the systematic investigation and evaluation. These questions help structure the review process, ensuring it remains focused on the most important aspects of interest.

\subsection{Problematic}
Medical imaging faces several challenges that impact the quality and reliability of diagnostic information. Issues such as noise, artifacts, large amounts of data, high-dimensional data, geometric deformations, and low resolution are common in various imaging modalities \cite{dhawan2011medical}. These challenges are especially detrimental to image registration, a critical process in medical imaging where information from different image modalities (e.g. MRI, CT) must be aligned and integrated. High noise levels and artifacts can lead to inaccuracies in the alignment process, while the large volume and complexity of medical image data make traditional registration methods computationally intensive and error-prone. Geometric deformations further complicate accurate registration, leading to suboptimal alignment and potentially misleading clinical interpretations. Various solutions have been proposed to address these challenges, including image denoising \cite{saladi2017analysis,pandey2016anatomization,lyra2012improved,mcgivney2014svd,fang2015spatio,yang2018low, kopriva2023low}, image compression \cite{doukas2007region,wang1996medical,zukoski2006novel}, image reconstruction \cite{he2022fidelity,asslander2018low,peng2016accelerated,ulas2016spatio,xu2017dynamic}, matrix completion \cite{li2020rank,friedland2006algorithm}, and tensor decomposition \cite{liu2020low,yi2021joint,cheng2014curvilinear,jia2022nonconvex}. While these methods have been effective in overcoming some of these challenges by reducing noise and capturing the essential features, they are not sufficient on their own. When using the LORMA, local variations in medical images can be overlooked, leading to the loss of important anatomical details \cite{liu2019adaptive,lee2016llorma}. In addition, LORMA struggles to handle the large amounts of data and high dimensionality of medical images, limiting its effectiveness in preserving key information \cite{kishore2017literature,nguyen2009fast,achlioptas2007fast,chu2003structured,liberty2007randomized}. The LLORMA technique is particularly important in the medical domain because it overcomes these limitations by allowing for variations in rank at the local level. This is especially relevant in medical imaging, where subtle variations in tissue characteristics can be clinically significant. LLORMA enables a more accurate representation of complex medical images, thereby improving the performance of image-processing tasks such as registration. Due to these gaps in current methods, the motivation for this paper is six-fold:
\begin{itemize}
 \item To understand and critically evaluate the application of LORMA and LLORMA in medical imaging.
 \item To discuss and propose how LORMA or LLORMA can be applied to various medical data types.
 \item To identify and analyze the medical datasets use in the literature with LORMA and LLORMA.
 \item To review and critique the similarity measurement methods employed to assess similar patches in medical imaging.
 \item To propose future research directions and potential solutions for improving the application of LORMA and LLORMA in medical imaging.
 \item To discuss the applicability and feasibility of the proposed solutions in real-world medical image scenarios.
\end{itemize}
This investigation addresses the research gaps by comparing LORMA and LLORMA in the context of medical imaging, highlighting their strengths, weaknesses, and limitations, and proposing a way forward in this field.

\subsection{Contributions}
In this study, we initially selected 59 publications for review. After applying quality assessment criteria, we narrowed the selection to 49 publications. Of these, 26 applied LORMA techniques to medical data, while 33 applied LLORMA techniques. We discuss the various datasets used in LORMA or LLORMA applications. Additionally, we summarize the tasks performed on these datasets and suggest other potential machine-learning tasks. Notably, we highlight a shift in preference toward LLORMA in the medical field since 2015, demonstrating its potential and effectiveness in capturing complex structures in medical data compared to LORMA. 

We acknowledge the limitations of shallow similarity methods commonly used in this field and suggest exploring advanced deep learning models such as DeepLab \cite{chen2017deeplab}. We provide a detailed explanation of how DeepLab can measure similar patches and assess its feasibility. Furthermore, our study highlights the importance of applying LORMA to various types of healthcare data. We discuss LORMA's limitations when handling irregular data types and demonstrate its applicability to regular data with missing entries. We further examine the impact of inaccuracies in predicting missing values and their effects on LORMA's performance. We also address the influence of patch size and propose using a random search (RS) technique to determine the optimal patch size. To enhance the feasibility of this approach, we introduce a hybrid method that combines Bayesian optimization with RS.

To the best of our knowledge, no systematic literature review specifically addresses LORMA and/or LLORMA with a focus on medical data. Unlike previous work that primarily focuses on LORMA, this review systematically evaluates its applications across different imaging modalities, identifies challenges and proposes innovative solutions to enhance its applicability and effectiveness. By addressing a significant gap in the literature and tackling practical issues such as computational scalability and similarity measurement, this work presents forward-looking strategies to advance the state of the art in medical imaging research.

\subsection{Manuscript Organization}
The structure of this work is as follows: Section~\ref{sec:methodology} oultines the methodology used in this paper. Section~\ref{sec:LRMA} examines the application of LORMA to medical data, highlighting various methods proposed by researchers to enhance the quality of medical imaging datasets, including MRI, CT, microarray, and infrared imaging. Section~\ref{sec:LLRMA} reviews the literature on the application of LLORMA across different medical imaging modalities. Section~\ref{sec:dataset} presents the medical datasets used in LORMA and LLORMA. Section~\ref{sec:similarity} explores various similarity measurement algorithms for identifying similar patches. Section~\ref{sec:findings} discusses the results and limitations of applying LORMA to medical data. Finally, Section~\ref{sec:recommendations} proposes future research directions, evaluates the applicability and feasibility of the suggested solutions, and serves as the study's conclusion by summarizing the main findings.

\section{Methodology}{\label{sec:methodology}}
This systematic review examines LORMA and LLORMA to establish a comprehensive and rigorous approach for identifying, selecting, and evaluating relevant field studies. The methodology outlines a systematic process to ensure the review’s transparency, reliability, and replicability. This process includes formulating the research question, defining inclusion and exclusion criteria, systematically searching literature databases, reviewing and selecting studies based on predefined criteria, extracting data from the selected studies, and assessing their quality and potential biases.

In this study, we follow a methodology based on the framework proposed by \cite{uman2011systematic} for systematic reviews. First, we define a strategy for conducting a comprehensive literature search. This search strategy is then executed to identify publications that meet the predefined inclusion and exclusion criteria. Relevant data are extracted through a thorough review and synthesis of the selected publications. Finally, the extracted data are analyzed and interpreted to summarize the current state of research and propose new directions for further investigation.

\subsection{Formulating Review Questions}
The study is guided by a comprehensive set of research questions, which help define the research's scope, applicability, and feasibility. Our objective is to review the literature on the application of LORMA and LLORMA in medical imaging, evaluating their effectiveness across various medical imaging modalities such as MRI, CT, and ultrasound. Specifically, we assess their performance in tasks like image denoising, reconstruction, and matrix completion. Moreover, we identify the limitations, challenges and conditions under which these methods perform optimally. We also examine the impact of similarity measure techniques and patch sizes on performance. To achieve these objectives, we formulated the research questions (RQN)  in Table~\ref{RQNS} to guide our investigation.

\begin{table} 
    \caption{Research questions guiding the review of LORMA and LLORMA in medical imaging.}
    \centering
    \begin{tabular}{p{0.05\linewidth}  p{0.4\linewidth}  p{0.5\linewidth}}
    \hline 
      {\textbf{RQN$i$}} & {\textbf{Question}} & {\textbf{Description}}\\ 
      \hline\hline\\
      {RQN1} & {Has the medical community switched to applying LLORMA rather than LORMA?} & {To identify whether the medical community has shifted from using LORMA to LLORMA in medical imaging and the reasons behind this shift.}\\
      \\
      {RQN2} & {Is LLORMA more effective than LORMA when applied to medical images?} & {To explore the considerations and modifications needed when applying LLORMA across different imaging modalities (e.g., MRI, CT).}\\ 
      \\
      {RQN3} & {Are there specific considerations or modifications required when applying LLORMA to different modalities of medical data?} &To explore the considerations and modifications needed when applying LLORMA across different imaging modalities (e.g., MRI, CT).\\ 
      \\
      {RQN4} & {What data types can LLORMA be applied to in the medical domain?} & {To identify and categorize the various data types in medical imaging where LLORMA can be effectively applied.}\\ 
      \\
      {RQN5} & {What are the limitations of LLORMA, and under what conditions does it perform well?} & {To identify the limitations of LLORMA and the specific conditions under which it performs best.}\\ 
      \\
      {RQN6} &What are the most commonly used similarity measure techniques in the context of LLORMA, and how do they impact the overall performance of LLORMA techniques? & To review the most common similarity measurement techniques, such as Euclidean distance, and how they impact the performance of LLORMA.\\ 
      \\
      {RQN7} & How does the choice of the local neighborhood or patch size impact the approximation quality, and what are the trade-offs involved? & {To assess the impact of different patch sizes on the approximation quality of LLORMA and the trade-offs involved.}\\ 
      \\[1ex]
      \hline
    \end{tabular}
    \label{RQNS}
\end{table}
\subsection{Data Sources and Search Strategy}{\label{search}}
In this systematic review, we utilized three databases -- Scopus, Web of Science, and PubMed to collect publications using advanced search techniques. Keywords were identified to construct the search string for selecting studies that addressed the research questions. Boolean operators were used to combine and refine these strings. Publications were initially selected based on their title, keywords, and abstracts. The search details are shown in Table~\ref{strings}.

\subsection{Selection and Data Extraction}
The selection of studies for this systematic review followed predefined inclusion and exclusion criteria to ensure relevance and quality. Following the search outlined in Section~\ref{search}, all publications were imported into separate Excel sheets and sorted by keywords. Duplicates were removed using the VLOOKUP and COUNTIF functions. The COUNTIF function counts the number of cells within a specified range that meet a given condition. In our case, it was used to count the number of times a specific publication appeared within a sheet to check for duplicates. The VLOOKUP function, on the other hand, was used to look up a publication in one sheet and return the corresponding publication from another sheet within the same Excel workbook, verifying if a publication in Sheet1 also appeared in Sheet2.

\begin{figure}
    \centering
    \includegraphics[scale=0.35]{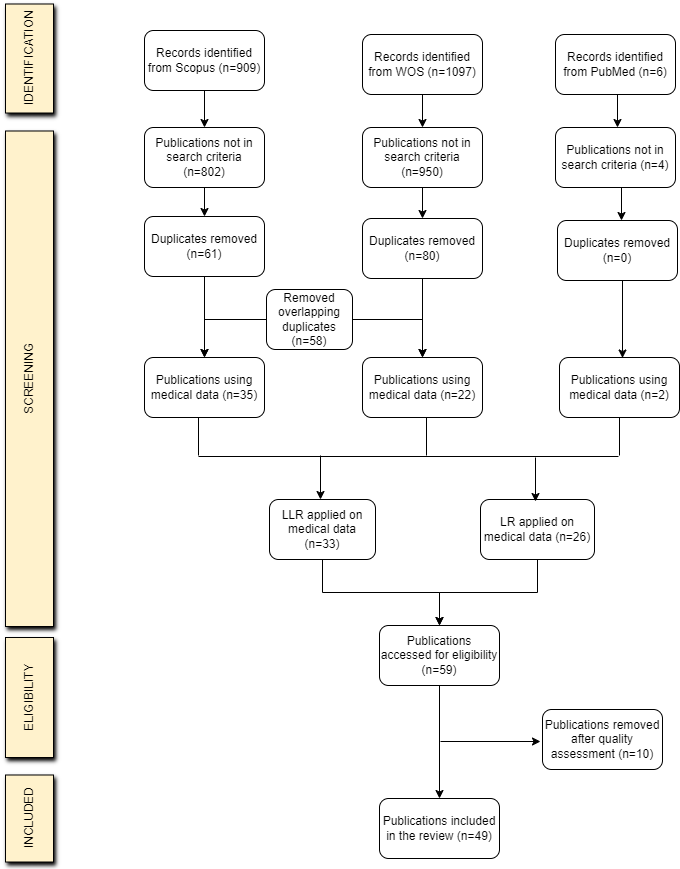}
    \caption{Prisma flow diagram depicting the study selection process for the systematic literature review.}
    \label{fig:prism}
\end{figure}

After duplicates were removed, relevant studies were selected based on the inclusion and exclusion criteria. The remaining studies were evaluated against the inclusion criteria, with any differences discussed and agreements reached. Fig.~\ref{fig:prism} illustrates the study selection procedure. A standardized form was developed for data extraction, which included key aspects such as study details (e.g., title, authors, publication year), LLORMA techniques (e.g., SVD, PCA), applications (e.g., image denoising, classification), main results, and conclusions. Two reviewers independently extracted data from the included studies.

\begin{table} 
    \caption{Search strings used in different databases.}
    \centering
    \begin{tabular}{p{0.1\linewidth}  p{0.5\linewidth}  p{0.1\linewidth}  p{0.1\linewidth}}
    \hline\hline 
      Database & Advanced search string & Search date & Results \\ 
      \hline\\
      Web of Science & ALL=((\q{low-rank approximation} OR \q{low-rank approximation} OR \q{local low-rank approximation} OR \q{local low-rank approximation}) AND (\q{image compression} OR \q{image filtering} OR \q{medical images})) & 02 Feb, 2024 & 1097 \\
      \\
      Scopus & TITLE-ABS-KEY(((\q{low-rank approximation}) OR (\q{low-rank approximation}) OR (\q{local low-rank approximation}) OR (\q{local low-rank approximation})) AND ((\q{image compression}) OR (\q{image filtering}) OR (\q{medical images}))) & 02 Feb, 2024 & 909\\ 
      \\
    PubMed & ((\q{low-rank approximation}) OR (\q{low-rank approximation}) OR (\q{local low-rank approximation}) OR (\q{local low-rank approximation})) AND ((\q{image compression}) OR (\q{image filtering}) OR (\q{medical images})) & 02 Feb, 2024 & 6\\ 
      Total & &  & 2012\\[1ex]
      \hline
    \end{tabular}
    \label{strings}
\end{table}
\subsection{Inclusion, Exclusion and Quality Assessment}
The study included publications that met specific inclusion criteria and underwent a quality assessment. 
\subsubsection{Inclusion and Exclusion Criteria}
The publications were collected from search engines (Scopus, Web of Science, and PubMed), and relevant studies were selected based on predefined inclusion and exclusion criteria.\begin{itemize}
    \item Inclusion criteria: Publications included in this systematic review met the requirement of applying matrix approximation and satisfied at least one of the following conditions:
\begin{itemize}
    \item[i.] The study uses a technique to compress, denoise, reconstruct or complete an image, among other applications.
    \item[ii.] The study discusses and applies LORMA or LLORMA to medical images.
\end{itemize}
\item  Exclusion criteria: Publications were excluded from this review if they met at least one of the following conditions:
\begin{itemize}
    \item [i.] The publication is not written in English.
    \item[ii.] The publication does not discuss LORMA or LLORMA.
    \item[iii.] The publication is a shortened version of another retrieved publication.
    \item[iv.] The publication is not peer-reviewed.
\end{itemize}
\end{itemize}
\subsubsection{Quality Assessment}
The quality of all selected publications was evaluated to assess the reliability and soundness of the research. Publications were reviewed using the following checklist:
\begin{itemize}
    \item Are the research objectives clearly defined? Is the statistical analysis appropriate for the study's research question and objectives?
    \item Does the data support the conclusions? Are the study's limitations and implications adequately discussed?
    \item  Are the references relevant, up-to-date, and correctly cited? Do other authors cite the study?
\end{itemize}
\section{LORMA  Applied to Medical Data}{\label{sec:LRMA}}
Researchers in medical imaging have proposed various methods to improve the quality of MRI, CT, microarray, and infrared imaging data. For instance, \cite{lyra2012improved} proposed using SVD-based LORMA approximation to reconstruct and denoise multi-dimensional MRI data. \cite{hu2023adaptive} introduced an extended Differential of Gaussian (DoG) filter and adaptive minimax-concave penalty for denoising MRI images. In contrast, \cite{mcgivney2014svd} used SVD to compress the size of the dictionary in the time domain for pattern recognition. \cite{liu2020low}, on the other hand, proposed a proximal operator for tensor nuclear norm approximation based on tensor-train rank-1 (TTr1) decomposition through SVD. 

In \cite{kopriva2023low}, the authors introduced low tensor-train rank and low multilinear rank approximations to address both despeckling and compression in tomography images, targeting a specified compression ratio. Regarding the calculation of optimal regularized inverse matrices with rank constraints, \cite{chung2017optimal} proposed an effective rank-update technique that decomposes the problem into smaller rank-related challenges. On the other hand, \cite{yi2021joint} suggested using multi-contrast Hankel tensor completion (MC-HTC) to leverage shared information in multi-contrast datasets, taking into account their highly correlated image structure, joint spatial support, and joint coil sensitivity for joint reconstruction. Meanwhile, \cite{friedland2006algorithm}, proposed an enhanced fixed-rank approximation algorithm (IFRAA) to address the limitations of the fixed-rank approximation algorithm (FRAA) in estimating missing values in a gene expression matrix. 

In the field of medical imaging, \cite{peng2016accelerated} proposed a method to improve the quality of fast T2 mapping, while \cite{asslander2018low} introduced a reconstruction framework for magnetic resonance fingerprinting (MRF). Another area of research has focused on spatiotemporal reconstruction for dynamic MRI, as demonstrated in the works of \cite{ulas2016spatio} and \cite{xu2017dynamic}. Additionally, \cite{fang2015spatio} proposed a spatiotemporal low-rank total variation method to recover arterial spin labeling MRI (ASL-MRI) data. Researchers have also explored LORMA methods for MRF dictionaries, as discussed by \cite{yang2018low}.

In other areas of medical imaging, various techniques have been proposed. For instance, \cite{mahoney2006tensor} extended matrix CUR decomposition to tensor-based datasets with distinguished modes. \cite{ding2018image} introduced an image-domain multi-material decomposition method for dual-energy computed tomography (DECT), aiming to suppress noise, reduce cross-contamination, and improve the accuracy of decomposed material images. \cite{li2020rank} proposed a method that factorizes an image matrix into a sum of rank-one matrices, where the rank is automatically estimated during convergence, eliminating the need for pre-specified rank information. Meanwhile, \cite{zhen2016descriptor} introduced a novel supervised descriptor learning (SDL) technique for multi-output regression, which constructs discriminative and compact feature representations to enhance multivariate estimate performance in medical images. The SDL algorithm extracts these representations using LORMA with supervised manifold regularization. Finally, \cite{he2022fidelity} proposed a multi-modality medical image fusion method for computer-aided diagnosis applications. This method suppresses noise in medical images and provides detailed information for disease diagnosis.

These studies highlight the significance of LORMA and regularization techniques in medical imaging. Table~\ref{LRSummaryTable} provides a summary of the aforementioned publications, detailing the low-rank methods applied to various medical datasets across different modalities. It includes information on the strengths and weaknesses of the papers using low-rank methods in various applications, along with references, year of publication, application, and the specific low-rank methods used.

\begin{landscape}
\pagestyle{empty}
\begin{table}
  \begin{center}
    \caption{Summary of the low-rank method, strengths and weaknesses of papers applying LORMA to medical datasets with different modalities.} 
    \label{tab:LRMA}
    \begin{tabular}{p{1.5cm} p{1cm} p{1.7cm} p{2.5cm} p{2cm} p{4cm} p{4cm}} 
        \hline
      \textbf{References} & \textbf{Year} &\textbf{Modality} & \textbf{Application}& \textbf{Low-rank method}& \textbf{Strengths}& \textbf{Weakness}\\
      \hline
      Mahoney et al. \cite{mahoney2006tensor}&2006&Hyperspectral imaging &Image compression and reconstruction& SVD &$\bullet$Captures the most essential underlying structures and patterns in the data for compression, reconstruction, and other tasks.&$\bullet$May not perform well for tensors with multiple equally important modes or similar properties.\\
      \hline
      Friedland et al. \cite{friedland2006algorithm}&2006&Genomics&Estimating missing values&SVD&$\bullet$Local-global algorithm exploiting local data similarity.&$\bullet$Computational complexity not extensively discussed; may be computationally intensive for large-scale microarray datasets.\\
      \hline
      Lyra et al. \cite{lyra2012improved}&2012&MRI&Image denoising& SVD&$\bullet$Estimates number of singular values using AIC, not just visual inspection.&$\bullet$Equally compresses the dataset.\\
\hline
Cheng et al. \cite{cheng2014curvilinear}&2014&X-ray&Tracking curvilinear structures& Tensor decomposition  &$\bullet$Uses tensor-based algorithm with model propagation for robust tracking.&$\bullet$Specifically designed for tracking deformable objects in X-ray images; may not apply directly to other types of images or objects.\\
\hline
Mcgivney et al. \cite{mcgivney2014svd}&2014&MRI&Image compression& SVD&$\bullet$Reduces computations without sacrificing signal-to-noise.&$\bullet$Applies equal weight to different singular values.\\
\hline
Fang et al. \cite{fang2015spatio}&2015&MRI&Image denoising&SVD &$\bullet$Uses joint spatial total variation and temporal low-rank regularization for improved SNR in ASL-MRI.&$\bullet$Due to the low signal-to-noise ratio, multiple acquisitions are required, leading to a longer scan time.\\
\label{LRSummaryTable}
   \end{tabular}
  \end{center}
\end{table}
\end{landscape}
\begin{landscape}
\pagestyle{empty}
\begin{table} 
\renewcommand\thetable{\ref{tab:LRMA}}
  \begin{center}
    \caption{Cont.}
     \begin{tabular}{p{1.5cm} p{1cm} p{1.5cm} p{2.5cm} p{2cm} p{4cm} p{4cm}} 
        \hline
      \textbf{References} & \textbf{Year} &\textbf{Modality}& \textbf{Application}& \textbf{Low-rank method}& \textbf{Strengths}& \textbf{Weakness}\\
      \hline
      Peng et al. \cite{peng2016accelerated}& 2016&MRI&Image reconstruction&SVD&$\bullet$Utilizes exponential parametric model in a relaxed manner to balance the trade-off between data consistency and the exponential structure.&$\bullet$Applies a uniform threshold to enforce Hankel low-rankness across all spatial locations.\\
      \hline
      Ulas et al. \cite{ulas2016spatio}&2016&DSC-MRI&Image reconstruction &Nuclear norm minimization &$\bullet$The suggested approach utilizes a reconstruction model that integrates penalties for dTV sparsity and nuclear norm simultaneously.&$\bullet$Nuclear norm minimization is sensitive to noise and outliers in the data.\\  
      \hline
      Xu et al. \cite{xu2017dynamic}&2017&MRI &Image reconstruction& RPCA &$\bullet$Employs a tighter nonconvex rank approximation, leading to improved image clarity and computational efficiency. &$\bullet$How effective the method is in different imaging modalities is not explored.\\
       \hline
      Chung et al. \cite{chung2017optimal}& 2017&MRI&Optimization&SVD&$\bullet$Breaks down the optimization issue into smaller subproblems of varying ranks and employs gradient-based techniques capable of leveraging linearity. &$\bullet$ Employing diverse noise levels in each problem or incorporating larger alterations in $k_j$ could lead to extended CPU times and/or elevated reconstruction errors in the update approach.\\
      \hline
      Assl{\"a}nder et al. \cite{asslander2018low}&2018&MRI&Reconstruction accuracy noise propagation&SVD&$\bullet$The LORMA approximation simplifies the computation by reducing the need for numerous Fourier transformations in the signal evolution.&$\bullet$The limitations and effectiveness of the proposed method in different imaging modalities are not explored.\\
      \hline
      Yang et al. \cite{yang2018low}&2018&MRI&Image compression&Randomized SVD&$\bullet$Decreases the memory needed to produce high-resolution MRF maps and enhances the overall speed of the compression process.&$\bullet$Randomized SVD may perform differently based on the specific properties of the MR fingerprinting matrix.\\
   \end{tabular}
  \end{center}
\end{table}
\end{landscape}
\begin{landscape}
\pagestyle{empty}
\begin{table} 
\renewcommand\thetable{\ref{tab:LRMA}}
  \begin{center}
        \caption{Cont.}
     \begin{tabular}{p{1.5cm} p{1cm} p{1.5cm} p{2.5cm} p{2cm} p{4cm} p{4cm}} 
        \hline
      \textbf{References} & \textbf{Year} &\textbf{Modality}& \textbf{Application}& \textbf{Low-rank method}& \textbf{Strengths}& \textbf{Weakness}\\
              \hline
      Ding et al. \cite{ding2018image}&2018&CT-scan&Image-domain multi-material decomposition&SVD&$\bullet$Enhances image resolution and accuracy through noise suppression and reduced cross-contamination.&$\bullet$Equal weighting to singular values in SVD.\\
      \hline
      Liu et al. \cite{liu2020low}&2020&MRI&Tensor completion&SVD&$\bullet$Utilizes tensor train rank-1 (TTr1) decomposition, enhancing decomposition accuracy.&$\bullet$Incorporates total variation as regularization, potentially unsuitable for datasets without significant local smoothness.\\
        \hline 
       Li et al. \cite{li2020rank}&2020&Ultrasound&Matrix completion &$l_p$-norm minimization &$\bullet$Robust to outliers, no need for rank or noise information.&$\bullet$Uniform weighting of singular values in $l_p$-norm minimization.\\ 
      \hline
      Yi et al. \cite{yi2021joint}&2021&MRI&Image reconstruction&HOSVD&$\bullet$Enables tensor low-rankness via higher-order SVD, improving reconstruction of multiple slices.&$\bullet$Iterative updates and higher-order SVD increase computational complexity.\\
      \hline
      He et al. \cite{he2022fidelity}&2022&MRI \quad PET&Reconstruction and image fusion &SVD &$\bullet$Utilizes rank coefficient optimization of LLORMA for multi-modality medical image reconstruction.&$\bullet$Equal weight application to different singular values in SVD.\\
      \hline
    Kopriva et al. \cite{kopriva2023low}&2023&Optical Coherence Tomography&Despeckling and compression& Tensor SVD&$\bullet$Simultaneously suppresses speckle and provides memory-efficient low-dimensional representations.&$\bullet$Evaluation on a limited dataset may not fully represent OCT image diversity in various clinical scenarios.\\
       \hline
Hu et al. \cite{hu2023adaptive}&2023&MRI&Image denoising&SVD&$\bullet$Improves the quality
of the image using the MCP model.& The computational complexity of the proposed method is not discussed.
         \end{tabular}
  \end{center}
\end{table}
\end{landscape}

\section{LLORMA Applied to Medical Data}{\label{sec:LLRMA}}
\label{sect:LLRMA}
LLORMA is a technique that is applied to medical data to effectively analyze and extract meaningful information while reducing computational complexity and noise. The term \textit{local low-rank} refers to finding low-rank sub-matrices within localized regions or areas of the data. Medical data often exhibit spatial or temporal correlations where neighboring regions exhibit similar patterns or structures. By applying LLORMA methods, we can exploit these local correlations and represent the data in smaller partial regions, where the rank of each localized region is lower than that of the entire dataset. This approach enables more efficient processing and analysis of medical data, especially in high-dimensional data. Applying LLORMA provides a robust framework for processing high-dimensional, spatially or temporally correlated data. By exploiting the inherent structure and redundancy within localized regions, LLORMA enables efficient processing, noise reduction, dimensionality reduction, and feature extraction. this, in turn, facilitates more accurate and insightful analysis in the medical domain. Medical data modalities, such as MRI, CT-scan, X-ray, ultra-sound, PET, multi-spectral imaging, and retinal imaging are commonly encountered in the medical field. This section reviews the literature on LLORMA's application to these different medical imaging modalities.\\

\subsection{LLORMA with MRI}
Recently, there has been increasing attention on developing efficient techniques to enhance the accuracy of clinical diagnoses by denoising 3D MRI data. Several methods for denoising 3D MRI images have been proposed in the literature. For instance, \cite{fu20163d} introduced a novel approach that fully exploits both local and non-local similarities in MRI using low-rank tensor approximation. Their method employs adaptive low-rank tensor approximation to effectively filter 3D MRI patch stacks, which is solved using the adaptive higher-order singular value thresholding (AHOSVT) algorithm. Similarly, \cite{chen2021novel} proposed a denoising method that integrates a non-local self-similarity technique with LORMA. Meanwhile, \cite{xia2017denoising} developed a technique based on LLORMA with weighted nuclear norm minimization (WNNM) to remove Rician noise from MR images. Since the tensor matricization method used by \cite{fu20163d} lacks flexibility, \cite{khaleel2018denoising} proposed an alternative method to denoise MR images affected by Rician noise using low-rank tensor approximation. Similarly, \cite{zhai2018weighted} conducted a study akin to \cite{xia2017denoising} but employed weighted Schatten p-norm as a minimization method.  

Further advancements include \cite{mandava2018accelerated}, who developed a reconstruction method for multi-contrast imaging and parameter mapping based on a union of local subspace constraints. \cite{lv2019denoising} extended the work of \cite{fu20163d} by integrating forward and inverse variance stabilizing transforms for the Rician distribution. Both \cite{fu20163d} and \cite{lv2019denoising} utilized denoising techniques based on the HOSVD approach, which applies a hard threshold function, limiting denoising performance. To address this limitation, \cite{wang2020modified} proposed modified HOSVD (MHOSVD), which optimizes denoising using a parameterized logarithmic nonconvex penalty function. Additionally, \cite{chen2021joint} introduced a denoising method that combines the DoG filter with non-local low-rank regularization. \cite{zhang2020image} proposed a structure-constrained LORMA (SLR) method that le leverages both the geometric information of local MRI image contents and the non-local self-similarity of global MRI structures. Their approach integrates two complementary regularization techniques: Kernel Wiener filtering (KWF) and low-rank regularization. On a different front, \cite{zhao2022joint} developed a joint denoising method for MR diffusion-weighted images (DWIs) using low-rank patch matrix approximation to exploit the natural redundancy in MRI data. Finally, \cite{he2022denoising} introduced a denoising technique based on weighted tensor nuclear norm minimization and balanced non-local patch tensors. 

\subsection{LLORMA with CT-scan}
Denoising CT images is a challenging task that requires sophisticated algorithms to reconstruct high-quality images from noisy inputs. Several methods have been proposed to tackle this issue using low-rank approximation and regularization techniques. For instance, \cite{lei2018denoising} developed a denoising algorithm based on low-rank sparse coding for CT image reconstruction. Their approach incorporates non-local self-similarity constraints and group-wise sparse coding noise regularization to minimize sparse coding errors, effectively improving image quality. similarly, \cite{sagheer2019denoising} introduced a denoising method that leverages both the global spatial correlation and local smoothness properties of CT images. Their technique involves extracting overlapping cubic patches, followed by vectorization. Similar patches are then grouped into blocks, forming a third-order tensor. The denoising process is performed using tensor singular value decomposition (t-SVD), while edge preservation and local smoothness are ensured through total variation (TV) regularization. In a related study, \cite{jia2022nonconvex} proposed a super-resolution model designed specifically for single 3D medical images. 

Recent advancements in CT image reconstruction have focused on integrating non-local low-rank regularization techniques with data-driven approaches to enhance image quality. For example, \cite{shen2021ct} proposed a novel non-local low-rank regularization and data-driven tight-frame CT image reconstruction model (NLR-DDTF). This model combines non-local low-rank matrix approximation (NLLORMA) and data-driven tight frame-based regularization, leveraging the strengths of both techniques to improve the accuracy of CT image reconstruction. More recently, \cite{chyophel2023low} introduced a denoising method that integrates a sparse 3D transform with Probabilistic Non-Local Means (PNLM). Their approach employs a sparse 3D transform with a hard-thresholding module to suppress noise in the transform coefficients, followed by collaborative Wiener filtering for further enhancement. Additionally, the PNLM component refines traditional non-local means by incorporating probabilistic weights that better capture patch similarity, making it more robust to complex noise patterns. These methods demonstrate the continued evolution of low-rank and non-local techniques in CT image denoising, improving both image clarity and diagnostic accuracy.

\subsection{LLORMA with X-Ray}
\cite{cheng2014curvilinear} propose a deformable tracking method based on tensors and model propagation to address the challenge of tracking anatomical structures and devices in X-ray images, particularly those prone to appearance changes and low visibility. The approach effectively solves the multi-dimensional assignment problem in medical imaging. In contrast, \cite{hariharan2019preliminary} introduced a denoising technique for 2D digital subtraction angiography (DSA) using a weighted local low-rank approach, facilitated by an advanced neurovascular replication system, to enhance image quality. Meanwhile, \cite{hariharan2018photon} proposed a photon recycling method for denoising ultra-low dose X-ray sequences, leveraging non-local self-similarity and weighted patch-matching to align similar image regions across frames. Their approach incorporates noise level-based LLORMA and weighted patch aggregation, resulting in improved denoised images. Both methods highlight the effectiveness of low-rank approximation techniques in enhancing the robustness and accuracy of medical imaging in complex environments.

\subsection{LLORMA with Ultra-sound}
\cite{sagheer2017ultrasound} proposed two methods for despeckling ultrasound images, from the speckled data. Despeckling is the process of removing speckle noise while maintaining edge preservation without sacrificing speckle reduction. The LLORMA is the first method used to despeckle ultrasound images (DLRA). This method converts multiplicative noise to additive noise through homomorphic filtering before applying weighted nuclear norm minimization (WNNM) since LLORMA methods for denoising an image are applicable to addictive noise. In 2019, \cite{sagheer2019despeckling} proposed an extension of the method. However, the direct extension of this approach disrupts the temporal correlation found in 3D US images, reducing despeckling performance. As a result, the novel method used tensor LLORMA to denoise 3D ultrasound images.

In a similar study, \cite{yang2021ultrasound} proposed a nonconvex LLORMA model based on non-local similar patch matrices to remove the speckle noise. The proposed method for ultrasound image restoration employs a nonconvex LLORMA model that incorporates the WNNM and data fidelity term.
\subsection{LLORMA with PET }
To effectively and efficiently apply tensor LLORMA, \cite{xie20193d} proposed a non-local tensor low-rank framework for dynamic PET reconstruction using a t-SVD-based technique. The proposed methodology integrates hidden data and enhances optimization efficiency by employing an expectation maximization (EM) method as a fidelity term and a TV constraint and a regularization method to extract local structures. Distributed optimization is utilized to jointly solve a Poisson PET model that incorporates these regularizations. Beyond spatiotemporal correlation, the Poisson model introduces a novel non-local low-rank tensor constraint that captures data correlations across multiple dimensions in dynamic PET. This model effectively recovers highly degraded data and enhances structures in low-active frames by leveraging temporal information between frames and non-local self-similarities within each frame. This approach increases structured sparsity for each image.

In contrast, \cite{yang2020pet} proposed a PET image denoising method based on LORMA that exploits self-similarity and low-rank of the images. The proposed method pre-filters the original noisy image in the first iteration to improve the accuracy of selecting similar patches. However, instead of directly grouping patches on the pre-filter image in the first iteration, the coordinates of the selected patches in the pre-filter image are recorded. These coordinates are used to group the patches in the noisy image. Next, similar patches are grouped as column vectors to form a group matrix, which is estimated using the LORMA method. All denoised patches are then aggregated to obtain an initially denoised image. To further enhance the results and recover lost information due to residual noise, an enhanced back projection technique is applied at each iteration. A final denoised image is obtained after several iterations. 

\subsection{LLORMA with multispectral imaging}
\cite{he2019segmenting} proposed a learning-based approach that takes advantage of the spectral and spatial characteristics of multispectral retinal images. The proposed method first represents each pixel's feature vector in each spectral MSI slice using a 2D spatial-spectral matrix. A generalized low-rank matrice approximation (GLRAM) framework is then employed to address the feature learning problem. The purpose of GLRAM is to construct low-dimensional, compact representations of a sequence of spatial-spectral matrices. The proposed method creates a 2D spatial-spectral matrix of local binary pattern features to accurately capture the spatial attributes of each spectrum image.

\subsection{LLORMA with retinal imaging}
\cite{ren2017drusen} proposed a method for drusen segmentation in retinal images using a supervised feature learning approach. The technique employs GLRAM and supervised manifold regularization to extract discriminative and concise descriptors from image patches obtained from retinal images. The learned features are closely associated with drusen, and may potentially be free of irrelevant information, enabling more accurate drusen identification against the background. In the final stage of the proposed model, a support vector machine classifier is used in the final stage of the proposed model to determine the presence of drusen in the pixels of test images.

\subsection{LLORMA with a Mixture of modalities}
Various methods for enhancing the quality and accuracy of medical images have been suggested in the existing literature. \cite{liu2015medical} proposed a method that combines LLORMA with an NNM constraint and block matching to fuse medical images and overcome distortion and information loss. \cite{zhong2016predict} proposed a kNN-regression method for predicting CT images from MRI data by identifying the nearest neighbors and employing supervised descriptor learning based on LORMA and manifold regularization. In contrast, \cite{yang2018predicting} developed a method for predicting pseudo-CT images from T1w and T2w MRI data using learned nonlinear local descriptors and feature matching. This method removes bias field artifacts from MR images using a bias correction algorithm. An intensity normalization technique additionally reduces the variance between MR images from different patients.

Another method proposed in \cite{liu2019medical} uses non-local self-similar redundancy and low-rank prior techniques to improve the resolution of low-resolution medical images and recover high-resolution images. The proposed method applies low-rank filtering to the group matrices to obtain pixel estimates. A weighted averaging approach is used to handle the issue of multiple estimates for each pixel. They performed a low-rank reconstruction and iteratively applied a sub-sampling consistency constraint to improve the method's performance. These methods enhance medical image analysis and visualization, improving the accuracy of medical diagnosis and treatment.

Table~\ref{tab:LLRMA} shows a comprehensive summary of similarity measures, low-rank methods, strengths, and weaknesses of papers applying the LLORMA technique to medical datasets with different modalities.
\begin{landscape}
\pagestyle{empty}
\begin{table}
  \begin{center}
    \caption{Summary of measure of similarity, low-rank method, strengths, and drawbacks of papers employing LLORMA on medical datasets with different modalities.} 
    \label{tab:LLRMA}
    \begin{tabular}{p{1.5cm} p{1cm}p{1.5cm} p{1.7cm} p{1.7cm} p{2cm} p{4cm} p{4cm}} 
        \hline
      \textbf{References} & \textbf{Year} & \textbf{Modality}& \textbf{Application}&\textbf{Measure of Similarity}& \textbf{Low-rank Method} &\textbf{Strengths}& \textbf{Weakness}\\
      \hline
      Liu et al. \cite{liu2015medical}& 2015& CT-scan\quad MRI\quad Ultrasound&Image fusion & Weighted Euclidean distance& SVD&$\bullet$Addresses fusion distortion and information loss.&$\bullet$Applies equal weights to different singular values for low-rank approximation.\\
      \hline
      Fu et al. \cite{fu20163d} & 2016 & MRI&Image denoising& KNN&Tensor approximation &$\bullet$Considers non-local spatial self-similarity and cross-slice correlation in 3D MR images &$\bullet$Lacks flexibility due to tensor matricization \cite{khaleel2018denoising}.\\
      \hline
         Zhong el al.\cite{zhong2016predict}& 2016&CT-scan\quad MR&Prediction of CT image from MRI &KNN  & GLRA  &$\bullet$Utilizes supervised descriptor learning based on LORMA and manifold regularization to optimize local MR image patch descriptors &$\bullet$KNN sensitivity to data noise and outliers.\\
         \hline
      Xia et al. \cite{xia2017denoising}& 2017&MRI&Image denoising&Block matching &Weighted nuclear norm minimization&$\bullet$Employs a flexible thresholding strategy.\newline $\bullet$Allows image attenuation without losing crucial features. & $\bullet$Inability to fully exploit correlation among different dimensions in high-dimensional MR data.\vspace{5pt} \newline $\bullet$Loss of high-frequency components such as edges and fine structures.\\
\hline
      Ren et al. \cite{ren2017drusen} &2017&Retinal imaging&Binary classification&Distance & GLRA&$\bullet$Diminishes image patch dimensionality while enhancing discriminative power.&$\bullet$Not always practical since it requires a large amount of labelled training data.\\
    \end{tabular}
  \end{center}
\end{table}
\end{landscape}
\begin{landscape}
\pagestyle{empty}
\begin{table} 
\renewcommand\thetable{\ref{tab:LLRMA}} 
  \begin{center}
    \caption{Cont.}
    \begin{tabular}{p{1.5cm} p{1cm} p{1.5cm} p{1.7cm} p{1.7cm} p{2cm}  p{4cm} p{4cm}} 
        \hline
      \textbf{References} & \textbf{Year} &textbf{Modality}& \textbf{Application}&\textbf{Measure of Similarity}& \textbf{Low-rank Method} &\textbf{Strengths}& \textbf{Weakness}\\
         \hline
      Sagheer et al. \cite{sagheer2017ultrasound}& 2017&Ultrasound &Image denoising&Block matching&  Weighted nuclear norm minimization&$\bullet$Includes a preprocessing stage considering statistical properties of ultrasound images. & $\bullet$Extending this method can compromise the temporal correlation in 3D ultrasound images, adversely affecting despeckling performance.\\
            \hline
Lei et al. \cite{lei2018denoising} & 2018&CT-scan &Image denoising& KNN & SVD&$\bullet$Utilizes group-wise sparse coding for noise minimization.\vspace{5pt} $\bullet$Employs Bayesian interpretation for adaptive parameter learning.&$\bullet$Robustness to different types or levels of noise is not extensively analyzed.\vspace{5pt} $\bullet$Computational requirements are not thoroughly discussed, potentially affecting feasibility.\\
\hline
Hariharan et al. \cite{hariharan2018photon}&2018&X-ray&Image denoising& Weighted patch matching& SVD&$\bullet$Enables parallel computing by denoising patches independently.&$\bullet$ Denoised images may exhibit mild blurring around instrument edges.\\
\hline
      Khaleel et al. \cite{khaleel2018denoising}&2018&MRI&Image denoising& Euclidean distance& t-SVD&$\bullet$Demonstrates flexibility and computational efficiency compared to traditional techniques.&$\bullet$The proposed method lacks generalizability for different medical image modalities.\\
      \hline
          Mandava et al. \cite{mandava2018accelerated}&2018&MRI&Reconstruction method for multi-contrast imaging&Subspace basis&SVD &$\bullet$Utilizes a union of local subspace constraints coupled with a sparsity-promoting penalty.&$\bullet$The reconstruction time for MOCCO-LS is also longer.\\
\hline
Zhai et al. \cite{zhai2018weighted}&2018&MRI&Image denoising&Block matching & Weighted Schatten p-norm minimization&$\bullet$The proposed method has a global optimum efficiently solvable by the generalized iterated shrinkage method when weights are in non-decreasing order. &$\bullet$Involves solving a computationally expensive weighted Schatten p-norm minimization problem.\\
   \end{tabular} 
  \end{center}
\end{table}
\end{landscape}
\begin{landscape}
\pagestyle{empty}
\begin{table} 
\renewcommand\thetable{\ref{tab:LLRMA}}
  \begin{center}
    \caption{Cont.}
    \begin{tabular}{p{1.5cm} p{1cm} p{1.5cm} p{1.7cm} p{1.7cm} p{2cm}  p{4cm} p{4cm}} 
        \hline
      \textbf{References} & \textbf{Year}& \textbf{Modality} & \textbf{Application}&\textbf{Measure of Similarity}& \textbf{Low-rank Method} &\textbf{Strengths}& \textbf{Weakness}\\
      \hline
 Yang et al. \cite{yang2018predicting}& 2018&CT-scan\newline MRI&Prediction of CT image from MRI &KNN  & GLRA &$\bullet$Utilizes a combination of dense scale-invariant feature transform descriptors and normalized raw patches for MR images, enhancing bone identification capability. &$\bullet$Lacks detailed discussion on the computational requirements, potentially impacting feasibility.\\
 \hline
      Sagheer et al. \cite{sagheer2019despeckling}& 2019&Ultrasound&Image denoising& $l_2$ distance & Tensor approximation &$\bullet$Effectively exploit spatial and temporal correlation in data for denoising. &$\bullet$ $l_2$ distance treats each pixel uniformly, neglecting contextual information and varying intensity significance across regions.\\
\hline
Liu et al. \cite{liu2019medical}&2019&CT-scan\newline MRI&Image denoising&Euclidean distance &SVD &$\bullet$Combines low-rank prior and non-local self-similar prior with iterative refinement for improved results.&$\bullet$Euclidean distance solely considers pixel intensities, overlooking inherent uncertainty or variability due to noise.\\
  \hline
Xie et al. \cite{xie20193d} &2019&PET&Image denoising and reconstruction& Euclidean distance &t-SVD & $\bullet$Spontaneously exploits inner temporal correlation without tracer information or model fitting. &$\bullet$Limited effectiveness for data with involuntary but noticeable motion. Not guaranteed to perform well in 3D PET reconstruction.\\
\hline
Sagheer et al. \cite{sagheer2019denoising}&2019&CT-scan &Image denoising& $l_2$ distance& Tensor approximation & $\bullet$Denoises images using tensor LORMA and tensor total variation technique. &$\bullet$Effectiveness depends on image characteristics and may not be suitable for all types of CT images.\\
   \end{tabular}
  \end{center}
\end{table}
\end{landscape}
\begin{landscape}
\pagestyle{empty}
\begin{table} 
\renewcommand\thetable{\ref{tab:LLRMA}}
  \begin{center}
    \caption{Cont.}
    \begin{tabular}{p{1.5cm} p{1cm} p{1.5cm} p{1.7cm} p{1.7cm}  p{2cm} p{4cm} p{4cm}} 
        \hline
      \textbf{References} & \textbf{Year} & \textbf{Modality} & \textbf{Application}&\textbf{Measure of Similarity}& \textbf{Low-rank Method} &\textbf{Strengths}& \textbf{Weakness}\\
            \hline
            Hariharan et al. \cite{hariharan2019preliminary}& 2019&X-ray &Image denoising&Euclidean distance & Weighted low-rank&$\bullet$Applies constrained low-rank approximations in both rows and columns.&$\bullet$The choice of a fixed patch size may not adequately consider anatomic variability.\\
        \hline
        Lv et al. \cite{lv2019denoising} & 2019 &MRI& Image denoising& KNN & Tensor approximation &$\bullet$ Exploits non-local and low-rank properties of the grouped image block.&$\bullet$Denoising methods based on HOSVD constrains the denoising performance \cite{wang2020modified}.\\
        \hline
      He et al. \cite{he2019segmenting}&2019&Multispectral imaging& Image segmentation& feature point matching& GLRA &$\bullet$Enables the learning of patterns closely related to their class labels for generating more discriminative representations.&$\bullet$The spatial-spectral features of other retinal diseases may vary from those of diabetic retinopathy, potentially affecting the efficacy of the proposed approach.\\
      \hline
     Yang et al. \cite{yang2020pet}& 2020 &PET& Image denoising& Euclidean distance& SVD &$\bullet$Utilizes an enhanced back-projection step to compensate for the loss of details, improving the denoising result &$\bullet$As Euclidean distance is not robust to noise in image data, a small amount of noise can significantly impact the distance calculation, resulting in inaccurate similarity measurements.\\
  \hline
 Wang et al. \cite{wang2020modified} &2020&MRI &Image denoising&Block matching&HOSVD &$\bullet$Uses nonconvex logarithmic regularization and non-local similarity to apply an adaptive multilinear tensor rank approximation approach.&$\bullet$The proposed method may not effectively eliminate noise in MR images with very low signal-to-noise ratios.\\
\hline
   Zhang et al. \cite{zhang2020image} &2020 &MRI&Image denoising&Block matching & SVD&$\bullet$Uses a structure-constrained LORMA model that exploits both local and non-local priors.&$\bullet$The method's effectiveness may depend on the specific characteristics of the images to be denoised.\\
\hline
Yang et al. \cite{yang2021ultrasound}&2021&Ultrasound &Image restoration &Euclidean distance& SVD&$\bullet$By incorporating the notion of fidelity, the method accounts for the unique characteristics of ultrasound images. &$\bullet$The choice of a fixed patch size may not sufficiently account for anatomic variability.
\end{tabular}
  \end{center}
\end{table}
\end{landscape}
\begin{landscape}
\pagestyle{empty}
\begin{table} 
\renewcommand\thetable{\ref{tab:LLRMA}}
  \begin{center}
    \caption{Cont.}
    \begin{tabular}{p{1.5cm} p{1cm} p{1.5cm} p{1.7cm} p{1.7cm}  p{2cm} p{4cm} p{4cm}} 
        \hline
      \textbf{References} & \textbf{Year}& \textbf{Modality}& \textbf{Application}&\textbf{Measure of Similarity}& \textbf{Low-rank Method} &\textbf{Strengths}& \textbf{Weakness}\\    
      \hline
      Chen et al. \cite{chen2021joint} & 2021 &MRI &Image denoising& Euclidean distance & SVD &$\bullet$High-frequency components are retrieved using DoG filter.&$\bullet$Euclidean distance is based on pixel intensities and does not consider the inherent uncertainty or variability caused by noise.\\
       \hline
      Chen et al. \cite{chen2021novel}&2021&MRI&Image denoising& Euclidean distance& SVD&$\bullet$Produces better approximation, especially when the noise level is high.&$\bullet$The method's sensitivity with respect to parameter settings, i.e., patch size, has not been thoroughly investigated.\\
      \hline
      Shen et al. \cite{shen2021ct} &2021 &CT-scan&Image reconstructing &Euclidean distance& SVD &$\bullet$Uses an asymmetric treatment for image reconstruction 
      and Radon domain inpainting &$\bullet$Euclidean distance is based on pixel intensities and does not consider the inherent uncertainty or variability caused by noise.\\
      \hline
      He et al. \cite{he2022denoising} &2022&MRI&Image denoising&K-means clustering& Tensor approximation&$\bullet$Constructs highly correlated 3D non-local patch tensors.\vspace{5pt} \newline$\bullet$Exploits the effectiveness of t-SVD using balanced 3D non-local patch tensors.&$\bullet$If the estimate of $\sigma_R$ is not accurate, the VST conversion may introduce residual errors or distortions that affect denoising performance.\\

      \hline
      Zhao et al. \cite{zhao2022joint} & 2022&MRI & Denoising diffusion-weighted images &Block matching &SVD &$\bullet$The proposed method has application to other clinical MRI protocols besides diffusion MRI.&$\bullet$If a large patch window size is used, the performance of the method may degrade.\\
      \hline
      Jia et al. \cite{jia2022nonconvex}&2022&CT-scan&High-resolution&Euclidean distance&Tensor decomposition&$\bullet$Avoids estimation bias caused by the traditional convex procedure.&$\bullet$Increased computational cost.\\
      \hline
      Chyophel et al. \cite{chyophel2023low}&2023&CT-scan&Image denoising&Distance threshold&SVD &$\bullet$The proposed method effectively reduces noise in LDCT images while preserving finer details and enhancing visual perception&$\bullet$The effectiveness of the method may be influenced by the assumptions made in noise modeling, particularly if the real noise deviates significantly from the modeled noise.
\end{tabular}
  \end{center}
\end{table}
\end{landscape}
\section{Medical Datasets Employed with LORMA and LLORMA}{\label{sec:dataset}}
Medical image datasets are collections of digital images obtained from various imaging technologies, such as PET, CT, and MRI. These datasets serve as valuable resources for training and developing machine learning algorithms and computer-aided diagnosis (CAD) systems in the medical field. This section provides a brief overview of several medical datasets that have been widely used in various machine learning applications.

One such dataset, frequently utilized in various studies \cite{fu20163d,xia2017denoising,zhai2018weighted,khaleel2018denoising,asslander2018low,lv2019denoising,liu2019medical,wang2020modified,he2022denoising,jia2022nonconvex}, is the synthetic 3D MRI dataset from the BrainWeb database. This dataset includes T1-weighted (T1w) data without noise, as well as T2-weighted (T2w) and proton density-weighted (PDw) data associated with a healthy brain. Furthermore, these studies used the real 3D MRI dataset from the Open Access Series of Imaging Studies (OASIS) database to validate the reliability of the proposed models. The OASIS dataset consists of 416 individuals aged 18 to 96 years, representing a cross-sectional sample. The participants include both males and females, all of whom are right-handed. Among them, 100 individuals aged sixty and older have been clinically diagnosed with very mild to moderate Alzheimer's disease (AD).

In addition to the aforementioned datasets, \cite{khaleel2018denoising} and \cite{wang2020modified} utilized the Auckland Cardiac MRI Atlas and the Multimodal Brain Tumor Image Segmentation Challenge (BraTS) 2013 databases, respectively. The Auckland Cardiac MRI Atlas dataset was acquired using a 1.5T SIEMENS Avanto MR device to obtain T1-weighted images, true FISP Cines, MR tagging, and contrast MRI. T1-weighted images and true FISP Cines were obtained through a retrospectively controlled steady-state free precession anatomic spin echo sequence, using a phased-array surface coil and ECG R-wave triggering. In contrast, the BraTS database provided pre-optimized MRI datasets through various image preprocessing methods, making them suitable for further applications such as CAD segmentations in simulation scenarios.

On the other hand, \cite{mahoney2006tensor} used hyperspectral image data comprising 59 data cubes from 59 biopsies, with each biopsy representing a different patient. Among these biopsies, there were 20 normal samples, 19 benign adenoma samples, and 20 malignant carcinoma colon samples. Each data cube contains a series of 128 grayscale images captured at a magnification of 400X. The images cover a frequency range of approximately 440 nm to 700 nm, with each image having dimensions of 495 pixels in width and 656 pixels in height.

\cite{yousefi2021diagnostic} utilized a breast cancer dataset comprising 208 participants enrolled in a breast screening study, categorized as either healthy (asymptomatic) or sick (symptomatic). The sick patients were diagnosed with breast cancer or non-cancerous conditions with symptoms using clinical breast examination (CBE) and mammography. The median age of the participants was 60 years, and they represented various ethnicities: 77 (37\%) were Caucasian, 57 (27.4\%) were African, 72 (34.6\%) were Pardo, 1 (0.5\%) was Mulatto, and 1 (0.5\%) was Indigenous. All participants underwent infrared imaging using a FLIR thermal camera (model SC620), with the acquired images having a spatial resolution of $640 \times 480$ pixels. 
\cite{hariharan2018photon} used 20 clinical sequences obtained from three different sites, involving seven patients and 500 images. These sequences were acquired using a significantly reduced radiation dose, termed ultra-low dose, which was set at 50\% of the standard low dose. The clinical ultra-low-dose sequences were captured during cardiac and renal electrophysiological (EP) procedures, with a matrix size of either $1024 \times 1024$ or $960 \times 960$ and a frame rate of 3 frames per second. \cite{cheng2014curvilinear} utilized X-ray fluoroscopic sequences from 17 clinical cases, with data acquired at a pixel size of $512 \times 512$ and a resolution of $0.4313~mm \times 0.3450~mm$. 

\cite{sagheer2019despeckling} employed three datasets, including real 3D ultrasound images and two distinct sets of simulated images. The real 3D ultrasound data consisted of images captured before and after tumor resection, while the simulated data included phantom images with dimensions of $128 \times 128 \times 128$ and images with a frequency of 20Hz, sized at $512 \times 512 \times 3$, respectively. \cite{shen2021ct} used a head image with smaller details, as well as head and brain images with more detailed structures, to evaluate the proposed method. 
 \cite{hariharan2019preliminary} analyzed X-ray sequences obtained at three different contrast agent concentration levels and four different dose levels relative to the standard dose. \cite{ren2017drusen} utilized the STARE and DRIVE databases. The STARE database contained 400 retinal fundus images, each with dimensions of $700 \times 605$ pixels, of which 63 were clinically diagnosed with drusen. The DRIVE database contained 40 RGB retinal images, each with dimensions of $768 \times 584$ pixels. \cite{he2022fidelity} used the brain atlas database, which included sections on normal brain conditions, cerebrovascular disease, neoplastic disease, degenerative disease, and infectious disease, with both MR and PET images. \cite{he2019segmenting} employed multispectral imaging (MSI) data consisting of 40 sequences representing unhealthy conditions and 10 sequences representing healthy conditions. These sequences captured images of the oculus dexter (OD) or oculus sinister (OS) from a cohort of 20 patients diagnosed with diabetic retinopathy (DR) and five healthy subjects. The images were in DICOM format with a bit depth of 16 and dimensions of $2048 \times 2048$ pixels. \cite{chen2021novel} and \cite{chen2021joint} used T2-weighted MR data acquired from a healthy volunteer using fast-field-echo sequences. The imaging parameters included a repetition time (TR) of 1050 ms, an echo time (TE) of 8.3 ms, a field of view of $280 \times 280mm^2$, and a flip angle of 111$^\circ$. \cite{asslander2018low} utilized a single slice of MR fingerprinting data of a healthy volunteer’s brain, acquired using a 3T Skyra scanner.  \cite{kopriva2023low} employed three-dimensional optical coherence tomography (OCT) images centered on the macula, obtained using the Topcon 3D OCT-1000 scanner. Each 3D OCT image consisted of 64 two-dimensional scans, each with dimensions of $480 \times 512$ pixels.

\begin{landscape}
\pagestyle{empty}
\begin{table}
\begin{center}
        \caption{This table provides an overview of the various datasets. Each dataset is described by a brief data description, the type of data, the tasks that can be performed on the dataset, the tasks performed, the links to access the dataset, and the publications that have used the data. The \q{x} symbol indicates that the possible tasks have not been performed on that dataset, while \q{$\checkmark$} indicates that the task has been performed on that particular dataset. Where dataset information was missing on a particular publication, the term \q{Not provided} was used for columns such as \q{Data description}, \q{Data source} and \q{Dataset}.}
    \label{tab:data}
    \begin{tabular}{p{2cm} p{4.8cm} p{1cm} p{3cm} p{2cm} p{2.5cm} p{2.7cm}}
        \hline
       
      \textbf{Dataset} &\textbf{Data description}& \textbf{Data type} & \textbf{Task}& \textbf{Performed task}&\textbf{Data source}& \textbf{Publications using this dataset}\\
    \hline
    Synthetic 3D BrainWeb MRI& Comprises T1w and T2w data without noise and a PDw data corresponding to the healthy brain. & Images &$\bullet$Image segmentation\newline $\bullet$Classification\newline$\bullet$Image denoising \newline$\bullet$Multi-modal fusion &x\newline x\newline \checkmark \newline x &\url{http://brainweb.bic.mni.mcgill.ca/brainweb}  &Fu et al. \cite{fu20163d}, Xia et al. \cite{xia2017denoising}, Zhai et al. \cite{zhai2018weighted}, Khaleel et al. \cite{khaleel2018denoising}, Asslander et al.\cite{asslander2018low}, Lv et al. \cite{lv2019denoising}, Liu et al. \cite{liu2019medical}, Wang et al. \cite{wang2020modified}, Chen et al. \cite{chen2021joint}, He et al.\cite{he2022denoising}\\
    \hline
    Real 3D MRI (OASIS)&Includes 416 cross-sectional subjects (aged 18-96) with 3-4 T1-weighted MRI scans per subject, including both right-handed men and women. It also features 100 subjects over 60 diagnosed with very mild to moderate Alzheimer's disease and a reliability subset of 20 non-demented subjects imaged within 90 days. & Images &$\bullet$Image segmentation\newline $\bullet$Classification\newline$\bullet$Image denoising \newline$\bullet$Multi-modal fusion &x\newline x\newline \checkmark \newline x & \url{http://www.oasis-brains.org/)}  &Fu et al. \cite{fu20163d}, Xia et al. \cite{xia2017denoising}, Zhai et al. \cite{zhai2018weighted}, Khaleel et al. \cite{khaleel2018denoising}, Lv et al. \cite{lv2019denoising}, Wang et al. \cite{wang2020modified}, He et al.\cite{he2022denoising}, Jia et al. \cite{jia2022nonconvex} Zhong et al. \cite{zhong2016predict}, Yang et al. \cite{yang2018predicting}, Peng et al. \cite{peng2016accelerated}\\
    \hline
     AMRG Cardiac MRI Atlas & The dataset comprises a labeled collection of MRI images depicting the heart of a healthy patient. &Images &$\bullet$Image segmentation\newline $\bullet$Classification\newline$\bullet$Image denoising \newline$\bullet$Multi-modal fusion &x\newline x\newline \checkmark \newline x & \url{https://www.cardiacatlas.org/amrg-cardiac-atlas/}& Khaleel et al. \cite{khaleel2018denoising}\\
     \hline
     BraTS & This dataset includes MRI datasets that have been preprocessed using various approaches to optimize them for further application simulations.&Images&$\bullet$Image segmentation\newline $\bullet$Classification\newline$\bullet$Image denoising \newline$\bullet$Multi-modal fusion &x\newline x\newline \checkmark \newline x &\url{https://www.smir.ch/BRATS/Start2013}&Wang et al. \cite{wang2020modified}\\
     \hline
      Real 3D CT & Consists of lung CT images in the DICOM format together with documentation of abnormalities by radiologists. &images&$\bullet$Image segmentation\newline $\bullet$Classification\newline$\bullet$Image denoising \newline$\bullet$Multi-modal fusion &x\newline x\newline \checkmark \newline x & \url{http://www.via.cornell.edu/databases}&Sagheer et al. \cite{sagheer2019denoising}\\
         \end{tabular} 
         \end{center}
         \end{table}
\end{landscape}
\begin{landscape}
\pagestyle{empty}
\begin{table}
\renewcommand\thetable{\ref{tab:data}}
  \begin{center}
    \caption{Cont.}
    \begin{tabular}{p{2cm} p{4.8cm} p{1cm} p{3cm} p{2cm} p{2.5cm} p{2.7cm}}
    \hline
      \textbf{Dataset} &\textbf{Data description}& \textbf{Data type} & \textbf{Task}& \textbf{Performed task}&\textbf{Data source}& \textbf{Publications using this dataset}\\
      \hline
      Low-dose CT & Consists of 4 images obtained from AAPM low-dose CT grand
challenge &images&$\bullet$Image segmentation\newline $\bullet$Classification\newline$\bullet$Image denoising \newline$\bullet$Multi-modal fusion &x\newline x\newline \checkmark \newline x &\url{http://www.aapm.org/GrandChallenge/LowDoseCT/} &Sagheer et al. \cite{sagheer2019denoising}\\
    \hline
    Simulated brain DWI &Data was generated with isotropic resolution = $2 mm$, 1 $b$
value = 0 image, and 30 diffusion directions with $b$
value = $1000$ and $2000 s/mm^2$&Images&$\bullet$Image segmentation\newline $\bullet$Classification\newline$\bullet$Image denoising \newline$\bullet$Multi-modal fusion &x\newline x\newline \checkmark \newline x  &Not provided&Zhao et al. \cite{zhao2022joint}\\
    \hline
In vivo, human brain DWI &Dataset was  acquired on a 3 Tesla Philips
scanner &Images&$\bullet$Image segmentation\newline $\bullet$Classification\newline$\bullet$Image denoising \newline$\bullet$Multi-modal fusion &x\newline x\newline \checkmark \newline x  &Not provided&Zhao et al. \cite{zhao2022joint}\\
        \hline
    T2 MR data &Obtained from a healthy volunteer with
fast-field-echo sequences: triceps surae of $1050/8.3 ms$, field of view of $280 \times 280mm^2$, flip angle of 111$^\circ$. &Images& $\bullet$Image segmentation\newline $\bullet$Classification\newline$\bullet$Image denoising \newline$\bullet$Multi-modal fusion &x\newline x\newline \checkmark \newline x  &Not provided&Chen et al. \cite{chen2021novel}, Chen et al. \cite{chen2021joint}\\
\hline
DSC-MRI brain perfusion sequence& Consists of brain perfusion sequence
of size $128 \times 128 \times 60$ with normalized intensities.&Images& $\bullet$Image segmentation\newline $\bullet$Classification\newline$\bullet$Image reconstruction \newline$\bullet$Image denoising &x\newline x\newline \checkmark \newline x&Not provided&Ulas et al. \cite{ulas2016spatio}\\

\hline
   Cardiac perfusion and cine& Consist of MRI images with dimensions $28 \times 128 \times 40$ for perfusion and  $28 \times 128 \times 24$ for cine &Images& $\bullet$Image segmentation\newline $\bullet$Classification\newline$\bullet$Image reconstruction \newline$\bullet$Image denoising &x\newline x\newline \checkmark \newline x& \url{https://cai2r.net/resources/}&Xu et al. \cite{xu2017dynamic}\\
   \hline
   In-vivo MRI data& Consist of an in-vivo MRI data with the dimensions $256\times256$&Images& $\bullet$Image reconstruction \newline$\bullet$Image denoising & \checkmark \newline x & \url{ http://shorty.usc.edu/class/591/fall04/}&Lyra et al. \cite{lyra2012improved}\\
   \hline
Multichannel 2D human brain datasets & Consist of T1-weighted,
T2-weighted, fluid-attenuated inversion recovery, and T1-weighted-inversion recovery.&Images& $\bullet$Feature extraction\newline $\bullet$Classification\newline$\bullet$Image reconstruction \newline$\bullet$Image denoising &x\newline x\newline \checkmark \newline x&Not provided&Yi et al. \cite{yi2021joint}\\
\hline
CT images& Consist of phantom, head and neck, and abdomen which were acquired using a Siemens CT scanner.&Images& $\bullet$Image segmentation\newline $\bullet$Classification\newline$\bullet$Image reconstruction \newline$\bullet$Image denoising &x\newline x\newline x \newline \checkmark &Not provided&Lei et al. \cite{lei2018denoising}\\
  \end{tabular}
  \end{center}
\end{table}
\end{landscape}
\begin{landscape}
\pagestyle{empty}
\begin{table}
\renewcommand\thetable{\ref{tab:data}}
  \begin{center}
    \caption{Cont.}
    \begin{tabular}{p{2cm} p{4.8cm} p{1cm} p{3cm} p{2cm} p{2.5cm} p{2.7cm}}
    \hline
      \textbf{Dataset} &\textbf{Data description}& \textbf{Data type} & \textbf{Task}& \textbf{Performed task}&\textbf{Data source}& \textbf{Publications using this dataset}\\
\hline
CT images&A head image with smaller details, a head image and a brain image with more details.&Images& $\bullet$Image segmentation\newline $\bullet$Classification\newline$\bullet$Image denoising \newline$\bullet$Image reconstruction &x\newline x\newline x \newline \checkmark  &Not provided&Shen et al. \cite{shen2021ct}\\
    \hline
Clinical data& Image sequences were acquired at 3 frames per second. There were 20
clinical sequences, that is, 7 patients and 500 clinical images. &Images/\newline video& $\bullet$Image segmentation\newline $\bullet$Classification\newline$\bullet$Image denoising \newline $\bullet$Object detection&x\newline x\newline \checkmark\newline x    &Not provided& Hariharan et al. \cite{hariharan2018photon}\\
\hline
    X-ray images& X-ray fluoroscopic sequences were obtained from 17 clinical cases. &Images/\newline video& $\bullet$Image segmentation\newline $\bullet$Classification\newline$\bullet$Image denoising \newline $\bullet$Tracking of deformable objects&x\newline x\newline x\newline \checkmark &Not provided& Cheng et al. \cite{cheng2014curvilinear}\\
\hline
X-ray images& X-ray sequences are obtained at three different concentration levels of the contrast agent and four different dose levels of the standard dose level. &Images/\newline video& $\bullet$Image segmentation\newline $\bullet$Classification\newline$\bullet$Image denoising \newline $\bullet$Dose optimization&x\newline \checkmark\newline x\newline x   &Not provided& Hariharan et al. \cite{hariharan2019preliminary}\\
\hline
MR fingerprinting data&A single slice MR fingerprinting data of a healthy volunteer’s brain that was acquired with a 3T Skyra scanner. &Images& $\bullet$Image segmentation\newline $\bullet$Classification\newline$\bullet$Image denoising \newline $\bullet$Image reconstruction&x\newline x\newline x\newline \checkmark &Not provided& Asslander et al. \cite{asslander2018low}\\
\hline
Real ultrasound data &Consists of ultrasound images with dimensions of $512\times 512$ &Images/\newline video&$\bullet$Image segmentation\newline $\bullet$Classification\newline$\bullet$Image despeckling \newline $\bullet$Image reconstruction&x\newline x\newline \checkmark\newline x &Not provided &Sagheer et al. \cite{sagheer2017ultrasound}, Yang et al. \cite{yang2021ultrasound}\\
\hline
Simulated ultrasound data &Consists of ultrasound images with dimensions of $512\times512$ &Images/\newline video& $\bullet$Image segmentation\newline $\bullet$Classification\newline$\bullet$Image despeckling \newline $\bullet$Image reconstruction&x\newline x\newline \checkmark\newline x &Not provided& Sagheer et al. \cite{sagheer2017ultrasound} Yang et al. \cite{yang2021ultrasound}\\
    \end{tabular}
  \end{center}
\end{table}
\end{landscape}
\begin{landscape}
\pagestyle{empty}
\begin{table}
\renewcommand\thetable{\ref{tab:data}}
  \begin{center}
    \caption{Cont.}
    \begin{tabular}{p{2cm} p{4.8cm} p{1cm} p{3cm} p{2cm} p{2.5cm} p{2.7cm}}
    \hline
      \textbf{Dataset} &\textbf{Data description}& \textbf{Data type} & \textbf{Task}& \textbf{Performed task}&\textbf{Data source}& \textbf{Publications using this dataset}\\
      \hline
    Real 3D ultrasound data & Consists of three-dimensional ultrasound images before and after tumor resection. &Images& $\bullet$Image segmentation\newline $\bullet$Classification\newline$\bullet$Image denoising \newline $\bullet$Image reconstruction&x\newline x\newline \checkmark\newline x  &\url{http://www.bic.mni.mcgill.ca/}& Sagheer et al. \cite{sagheer2019despeckling}\\
    \hline
Simulated 3D ultrasound data & consists of phantom images corresponding to a size of $128 \times 128 \times 128$. &Images& $\bullet$Image segmentation\newline $\bullet$Classification\newline$\bullet$Image denoising \newline $\bullet$Image reconstruction&x\newline x\newline \checkmark\newline x  &\url{https://in.mathworks.com/matlabcentral/}& Sagheer et al. \cite{sagheer2019despeckling}\\
\hline
Simulated 3D ultrasound data & consists of simulated 3D images at a frequency of 20Hz of size $512 \times 512 \times 3$. &Images& $\bullet$Image segmentation\newline $\bullet$Classification\newline$\bullet$Image denoising \newline $\bullet$Image reconstruction&x\newline x\newline \checkmark\newline x  &\url{http://splab.cz/en/research/zpracovani-medicinskych-signalu/databaze/artery}& Sagheer et al. \cite{sagheer2019despeckling}\\
\hline
Zubal head phantoms and Real Cardiac data& The data are scanned over 60 min by a Hamamatsu SHR-22000.&Images&$\bullet$Image segmentation\newline $\bullet$Classification\newline$\bullet$Image reconstruction \newline $\bullet$Image denoising&x\newline x\newline \checkmark\newline x  &Not provided&Xie et al. \cite{xie20193d}\\
\hline
CT lung and cardiac data& Data was obtained from the National Institute of Healths Lung Imaging Database Consortium&Images&$\bullet$Image segmentation\newline $\bullet$Classification\newline$\bullet$Image resolution enhancement \newline $\bullet$Image denoising&x\newline x\newline \checkmark\newline x &Not provided&Liu et al. \cite{liu2019medical}\\
\hline
CT images & Data was acquired on a GE Medical Systems Light
Speed Pro 16 scanner&Images& $\bullet$Image segmentation\newline $\bullet$Classification\newline$\bullet$CT image prediction \newline $\bullet$Image denoising&x\newline x\newline \checkmark\newline x&Not provided& Zhong et al. \cite{zhong2016predict}, Yang et al. \cite{yang2018predicting}\\
\hline 
MSI images & Consist of 40 sequences indicative of unhealthy conditions and ten sequences representing healthy conditions. The images were in DICOM format with a bit depth of 16, each having dimensions of $2048 \times 2048$.&Images&  $\bullet$DR lesion segmentation\newline $\bullet$Classification\newline$\bullet$Image denoising \newline $\bullet$Image reconstruction&\checkmark\newline x\newline x\newline x &Not provided& He et al. \cite{he2019segmenting}\\
        \end{tabular}
  \end{center}
\end{table}
\end{landscape}
\begin{landscape}
\pagestyle{empty}
\begin{table}
\renewcommand\thetable{\ref{tab:data}}
  \begin{center}
    \caption{Cont.}
    \begin{tabular}{p{2cm} p{4.8cm} p{1cm} p{3cm} p{2cm} p{2.5cm} p{2.7cm}}
    \hline
      \textbf{Dataset} &\textbf{Data description}& \textbf{Data type} & \textbf{Task}& \textbf{Performed task}&\textbf{Data source}& \textbf{Publications using this dataset}\\
    \hline
STARE database & Contains 400 retinal fundus images of the retina, each image having a size of $700 \times 605$. Sixty-three images were clinically found to contain drusen. &Images&$\bullet$Image segmentation\newline $\bullet$Classification\newline$\bullet$ Optic disc localization \newline$\bullet$Generic
lesion detection &x\newline \checkmark\newline x \newline x   &\url{http://cecas.clemson.edu/~ahoover/stare/}& Ren et al. \cite{ren2017drusen}\\
\hline
DRIVE database& Contains 40 RGB color retinal images; each image is $768\times584$. &Images&$\bullet$Image segmentation\newline $\bullet$Classification\newline$\bullet$ Optic disc localization \newline$\bullet$Generic
lesion detection &x\newline \checkmark\newline x \newline x   &\url{https://www.isi.uu.nl/Research/Databases/}& Ren et al. \cite{ren2017drusen}\\
\hline
Hyperspectral image& Comprises 59 sets of data cubes obtained from 59 biopsies, with each biopsy representing a different patient. Among these biopsies, there are 20 normal samples, 19 samples of benign adenoma, and 20 samples of malignant carcinoma colon. &Images&$\bullet$Target detection\newline $\bullet$Classification\newline$\bullet$ Anomaly detection \newline$\bullet$Image compression&x\newline \checkmark\newline x \newline \checkmark&Not provided& Mahoney et al. \cite{mahoney2006tensor}\\
\hline
Brain atlas& Contains the sections on normal brain, cerebrovascular disease,
neoplastic disease, degenerative disease, and infectious disease
of MR and PET images&Images& $\bullet$Image segmentation\newline $\bullet$Classification\newline$\bullet$Image denoising \newline$\bullet$Multi-modal fusion &x\newline x\newline x \newline \checkmark &\url{http://www.med.harvard.edu/AANL
IB/home.html}&He et al.~\cite{he2022fidelity}\\
\hline
Infrared images&A total of 208 individuals were included in the study for breast screening, consisting of both healthy individuals without symptoms and individuals with symptoms.&Images& $\bullet$Image segmentation\newline $\bullet$Classification\newline$\bullet$Image denoising \newline$\bullet$Lesion Detection &x\newline \checkmark\newline x \newline x &Not provided&Yousefi et al. \cite{yousefi2021diagnostic}\\
\hline
OCT&OCT images of normal eyes acquired with the Topcon 3D OCT-1000 scanner&images& $\bullet$Image segmentation\newline $\bullet$image compression\newline$\bullet$Image despeckling \newline$\bullet$Classification &x\newline \checkmark\newline \checkmark \newline x &\url{https://pan.baidu.com/share/init?surl=edkG7k8W3Wkjhq8vffCYng}&Kopriva et al. \cite{kopriva2023low}      
\end{tabular}
  \end{center}
\end{table}
\end{landscape}
\section{Similarity Measure Algorithms}{\label{sec:similarity}}
\label{sec_1}
Various similarity measures can be utilized for medical data analysis and pattern recognition in general. The concept of a similarity measure plays a crucial role in various fields of study, such as data mining, ML, and information retrieval. Similarity measures are used to quantify the degree of relatedness between two data points based on some specified criteria. In medical data analysis, data embeddings can be used to represent relationships between medical conditions and symptoms. For instance, an embedding could be created to show that the symptom of chest pain is related to the medical condition of myocardial infarction, while shortness of breath is related to heart failure. This aligns with the medical notion of disease categorization and conceptual hierarchies, which can aid in diagnosis, treatment planning, and patient outcomes. By using latent representations in medical data analysis, healthcare professionals can identify patterns, cluster similar conditions together, and make informed decisions. Thus, deriving similarities, whether among physical concepts or abstract ones, is very important in pattern recognition. 
Similarity measures are also utilized in LLORMA to identify similar patches within a medical image when the LORMA technique is applied for tasks such as denoising, dimensionality reduction, compression, and feature extraction.

This section provides a detailed discussion of five different similarity measurement algorithms used in the LLORMA methods applied to medical imaging, as discussed in Section~\ref{sect:LLRMA}. These similarity measures include the K-nearest neighbor (KNN) algorithm, K-means clustering, Euclidean distance and weighted Euclidean distance, weighted patch matching, and block matching using the $l_1$ norm, also known as Manhattan distance. We will present an in-depth explanation of each method, highlighting its advantages, limitations, and specific use cases. By understanding the characteristics and performance of these similarity measures, we aim to help researchers and practitioners select the most appropriate measure for their LLORMA applications.

\subsection{Distance Measures}
\label{sec:distance_measures}
\begin{figure}
    \centering
    \includegraphics[width=1.0\textwidth]{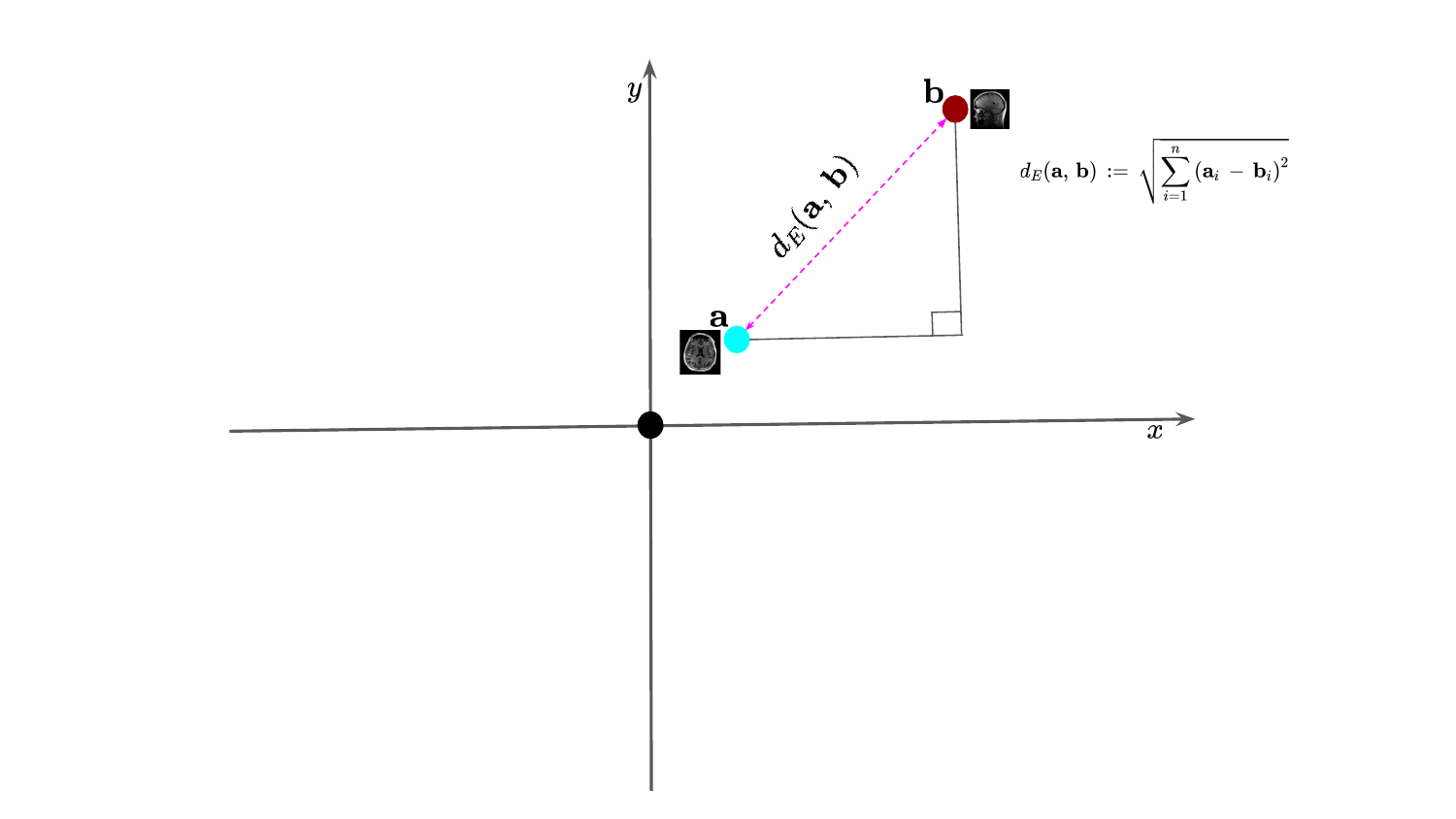}
    \caption{The Euclidean distance is a commonly used measure of similarity between two points in a Euclidean space. It is defined as the square root of the sum of the squared differences between the corresponding coordinates of the two points as shown in this illustration where $\mathbf{a}$ and  $\mathbf{b}$ are two vectors in $\mathbb{R}^2$. These can be, for example, representations of two MRI scans as depicted in this figure.  However, this notion generalizes to $\mathbb{R}^n$.}
    \label{fig:distances}
\end{figure}

\begin{figure}
    \centering
    \includegraphics[width=1.0\textwidth]{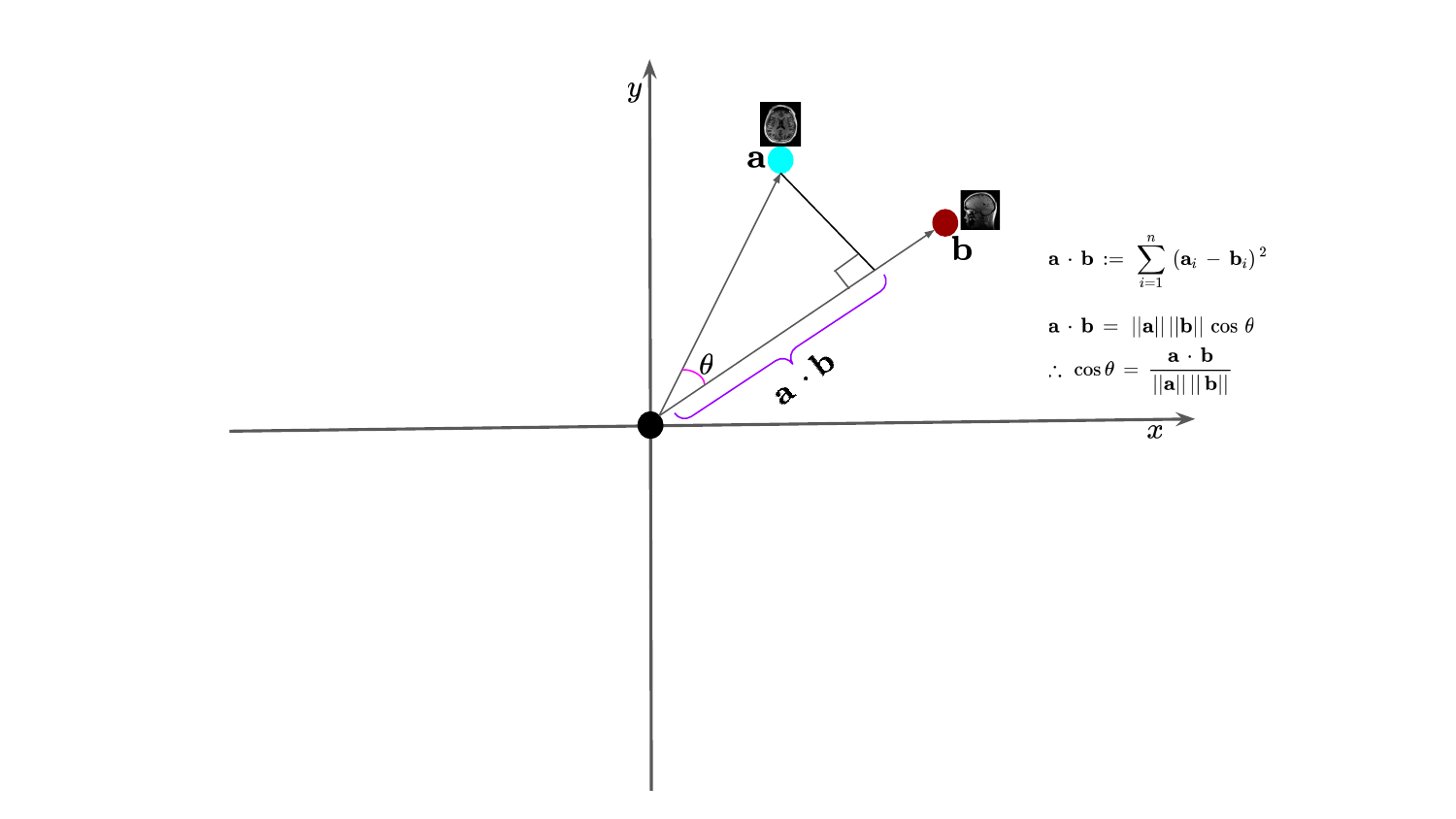}
    \caption{The dot product between two vectors ($\mathbf{a}$ and $\mathbf{b}$), is defined as the sum of the products of their corresponding components. As shown in the figure, it is related to the Euclidean distance.}
    \label{fig:distances2}
\end{figure}

In pattern recognition and machine learning, a distance measure, as defined in Eq.~\ref{eq:dist}, quantifies the similarity or dissimilarity between two objects based on their features or attributes~\cite{bishop2006pattern,goodfellow2016deep, roberts2022principles}. These measures, often referred to as distance functions or metrics in mathematics, are used to compare data points (e.g., vectors in a vector space) and are fundamental to many learning algorithms. A distance function assigns a numerical value to the relationship between two objects, with smaller values indicating greater similarity and larger values indicating greater dissimilarity. Fig.~\ref{fig:distances} and Fig.\ref{fig:distances2} provide geometric illustrations of this concept.
\begin{equation}
    d: X \times Y \to \mathbb{R}_{\geq 0}.
    \label{eq:dist}
\end{equation}

There are numerous types of distance measures, and an entire mathematical field called Measure Theory is dedicated to studying them, along with related concepts such as volume, probability, and Lebesgue measure. Some commonly used distance measures include Euclidean distance, Manhattan distance, and Hamming distance~\cite{gray2009probability,cohn2013measure,halmos2013measure,hamming1950error}. Each of these measures has its own mathematical formulation and is suitable for different types of data, depending on the specific features or attributes being compared. Selecting an appropriate distance measure is crucial for ensuring the accuracy and effectiveness of learning algorithms across various applications. For a function to be considered a valid distance measure, it must satisfy four key properties:
\begin{itemize}
    \item Non-negativity: If $d(\mathbf{a} , \mathbf{b})$ is the distance between points $\mathbf{a} \in X$ and $ \mathbf{b} \in Y$, then $d(\mathbf{a} , \mathbf{b}) \geq 0$.
    \item Identity: If $ \mathbf{a} \in X$ and $ \mathbf{b} \in Y$ are identical points, then $d(\mathbf{a} , \mathbf{b}) = 0$.
    \item Symmetry: If $ \mathbf{a} \in X$ and $ \mathbf{b} \in Y$ are any two points, then $d(\mathbf{a}, \mathbf{b}) = d(\mathbf{b},\mathbf{a})$.
    \item Triangle inequality: If $ \mathbf{a} \in X$, $ \mathbf{b}\in Y$, and $ \mathbf{c} \in Z$ are any three points, then $d(\mathbf{a} , \mathbf{b}) + d(\mathbf{b} , \mathbf{c}) \geq d(\mathbf{a} , \mathbf{c})$.
\end{itemize}
These four properties ensure that a \textit{true} distance measures a valid mathematical function and produces meaningful and consistent results. Distance measures satisfying these properties are widely used in various applications, including clustering, classification, and pattern recognition.

The Euclidean distance (often referred to as the $l_2$ norm) and cosine similarity are two commonly used measures for evaluating similarity between vectors in tensor analysis. quantifies the straight-line distance between two points in an $n$-dimensional space, whereas cosine similarity measures the cosine of the angle between two vectors, reflecting their directional alignment rather than magnitude. The relationship between these two measures can be understood through their underlying mathematical connection, which stems from the geometric interpretation of vectors. Specifically, given two vectors $\mathbf{a}$ and $\mathbf{b}$, the Euclidean distance $d_E(\mathbf{a},\mathbf{b})$ is computed as the square root of the sum of the squared differences between their corresponding coordinates, expressed as: 

\begin{flalign}
    d_E(\mathbf{a},\mathbf{b})&=\sqrt{\sum_{i=1}^n(\mathbf{a}_i-\mathbf{b}_i)^2}\\
    &= \sqrt{(\mathbf{a} - \mathbf{b}) \cdot (\mathbf{a} - \mathbf{b})}.
    \label{eq:euclidian}
\end{flalign}
On the other hand, cosine similarity is given by the dot product of the two vectors, normalized by the product of their magnitudes, i.e.,
\begin{equation}
    \cos(\theta)=\frac{\mathbf{a}\cdot\mathbf{b}}{\lVert \mathbf{a} \rVert \lVert \mathbf{b} \rVert}.
\end{equation}
The cosine similarity can be related to the Euclidean distance by the identity 
\begin{flalign}
    \cos(\theta)&=\frac{\mathbf{a}\cdot\mathbf{b}}{\lVert \mathbf{a} \rVert \lVert \mathbf{b} \rVert}\\
    &=\frac{\sum_{i=1}^na_ib_i}{\sqrt{\sum_{i=1}^na_i^2}\sqrt{\sum_{i=1}^nb_i^2}},
\end{flalign}
which shows that the cosine similarity is a normalized version of the dot product of the vectors. Therefore, while Euclidean distance measures the absolute distance between two vectors, cosine similarity measures their relative orientation. These measures are often used in combination with ML algorithms such as k-means clustering, nearest neighbor classification, and LLORMA where they play complementary roles in measuring similarity and distance.

\subsection{Euclidean Distance and Weighted Euclidean  Distance}
The Euclidean distance can be used to measure the similarity between two vectors with respect to their magnitude and direction. However, the weighted Euclidean distance (WED) is a modification of Euclidean distance, where each coordinate of the vectors is multiplied by a weight factor. These weights are used to adjust the contribution of each coordinate to the distance measure. This is computed as:

\begin{equation}
d_w(\mathbf{a},\mathbf{b}) = \sqrt{\sum_{i=1}^n \mathbf{w}_i(\mathbf{a}_i - \mathbf{b}_i)^2},
\label{eq:weighted_euclidian_dist}
\end{equation}
where $\mathbf{a}$ and $\mathbf{b}$ are two vectors of dimension $n$ and $\mathbf{w}_i$ is the weight factor associated with the $i$th coordinate.

The WED is commonly used in fields such as data mining and image processing, where the data may have a non-uniform distribution or varying importance of the features. The weights can be used to emphasize the importance of certain features or to reduce the impact of irrelevant or noisy features. Both Euclidean distance and WED are useful measures of similarity between vectors and can be applied to a variety of fields, including ML, pattern recognition, and data analysis.
\subsection{K-Nearest Neighbors}
K-Nearest Neighbors (K-NN) is a widely used ML algorithm in data analysis for classification and regression problems~\cite{khalid2011mri,hu2016distance,mack1981local}. The K-NN algorithm is a non-parametric and lazy (instance-based) learning method that does not make any assumptions about the underlying data distribution. The basic idea behind K-NN is to find the K-nearest neighbors to a given data point (which we shall call a \textit{query}) based on a distance metric and use the labels of those neighbors to classify or predict the label of the given data point as shown in Algorithm~\ref{alg:knn}. In the case of classification, the majority vote of the K-nearest neighbors is taken as the predicted label of the data point, while in regression, the average of the K-nearest neighbors' labels is taken as the predicted value.

To apply the K-NN algorithm, we first need to choose an appropriate distance metric. In medical data analysis, the Euclidean distance and WED are commonly used as distance metrics.

\begin{algorithm}[]
\caption{K-Nearest Neighbors}
\label{alg:knn}
\begin{algorithmic}[1]
\Require Dataset $X = \{(\mathbf{x}_1, y_1), (\mathbf{x}_2, y_2), \ldots, (\mathbf{x}_n, y_n)\}$, query point $\mathbf{x}$, number of neighbors $K$, distance metric $d(x, y)$

\State Initialize distances array $D = []$

\For{$i=1$ \textbf{to} $n$}
    \State Compute distance $d_i = d(\mathbf{x}, \mathbf{x}_i)$
    \State Append distance $D.\text{append}(d_i)$

\EndFor
\State Sort $D$ in ascending order
\State Get the first $K$ distances in $D$: $D_K$
\State Get the indices of $D_K$: $I_K$
\If{classification}
 \State $y = \text{mode}( {y_i \mid  i \in I_K})$
\ElsIf{regression}
    \State $y = \text{mean}\left( {y_i \mid i \in I_K}\right)$
\EndIf

\Return Predicted label $y$
\end{algorithmic}
\end{algorithm}

\begin{figure}
    \centering
    \includegraphics[width=\textwidth]{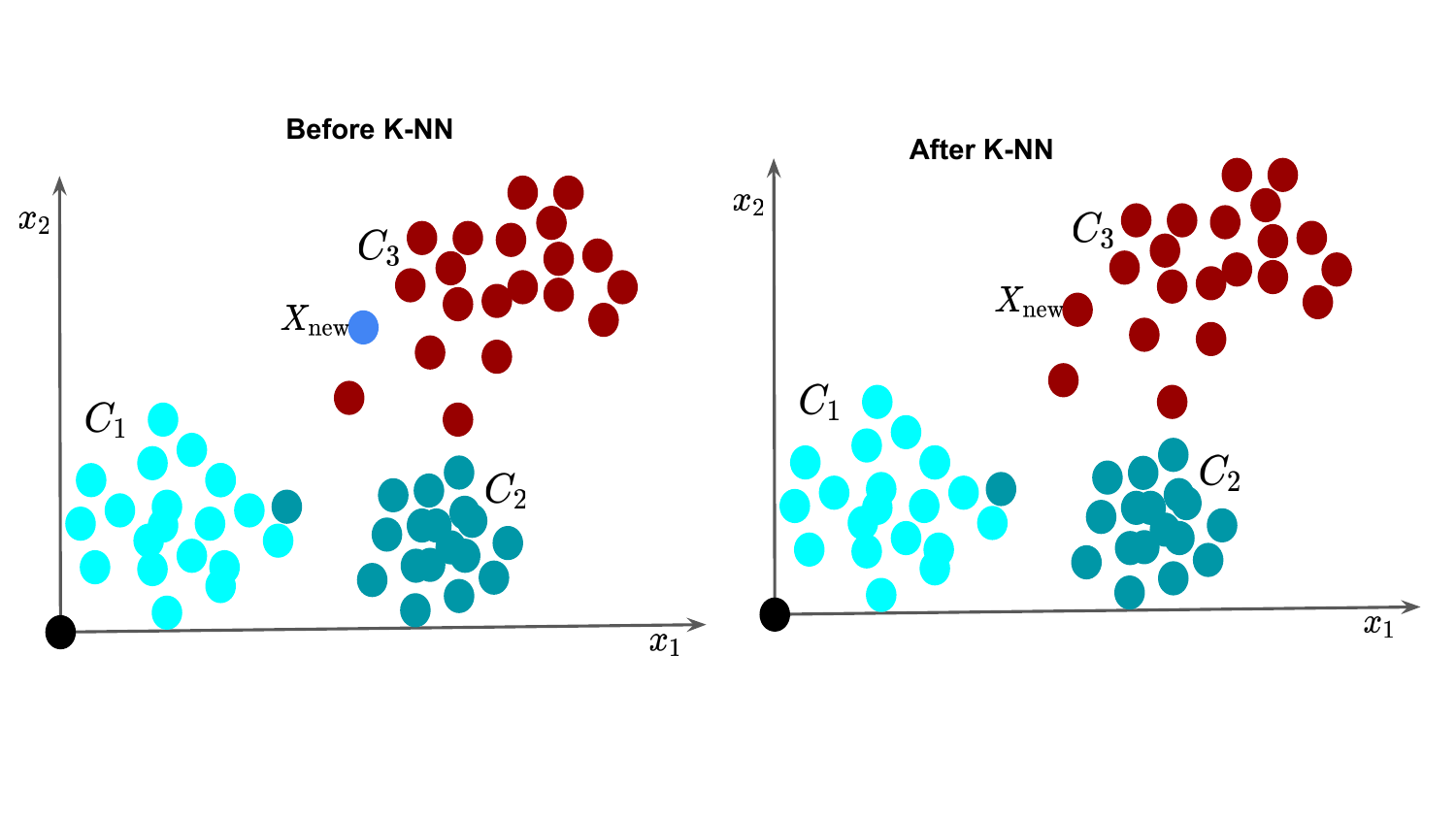}
    \caption{(Left) A 2D feature visualization of a dataset that comprises three classes ($C_1, C2, C_3$). The colors indicate the class labels. With a new datapoint $X_\mathrm{new}$, the objective of the K-NN algorithm is to find the class $C_i$ to which the datapoint can be classified based on its $K$ neighbors labels. (Right) After computing the distance between the query $X_\mathrm{new}$ and all the samples in the dataset, it has been classified into class $C_3$, which comprises a majority of the new sample neighbors.}
    \label{fig:KNN_Illustration}
\end{figure}

However, in some cases, not all features in the data may be equally important. For instance, in medical data analysis, some features may be more clinically relevant than others. In such cases, K-NN uses a WED, which assigns different weights to each feature based on its importance.  The weights can be assigned based on domain knowledge or learned from the data using techniques such as principal component analysis (PCA). We have shown a visual illustration of the K-NN algorithm in Fig.~\ref{fig:KNN_Illustration}.
In the K-NN algorithm, the choice of $K$ is a hyperparameter that needs to be tuned since it can lead to overfitting or underfitting depending on the value. Methods such as cross-validation or grid search can be used to obtain the optimal value of $K$. The choice of distance metric, such as Euclidean distance and WED, and the value of $K$ need to be carefully chosen based on the specific problem and data at hand.

\subsection{K-Means Clustering}
$K$-means clustering is a commonly used technique in various domains, including medical data analysis~\cite{bi2020computer}. The goal of $K$-means clustering is to group similar data points together based on their features. As shown in Algorithm~\ref{alg:kmeans}, an initial guess of $K$ centroids of the clusters is given and then each point is iteratively assigned to its nearest centroid based on a distance metric such as Euclidean distance, and the centroid is updated based on the mean of the assigned data points within a given cluster. 
\begin{algorithm}
\caption{K-Means Clustering}
\label{alg:kmeans}

\begin{algorithmic}[1]

\Require Number of clusters $K$, dataset $(\mathbf{x}_1, \mathbf{x}_2, \ldots, \mathbf{x}_n)$

\State Initialize centroids $\boldsymbol{\mu}_1, \boldsymbol{\mu}_2, \ldots, \boldsymbol{\mu}_K$

\Repeat

\For{$i=1$ {\bfseries to} $n$}

\State Compute distances: $d_j = \lVert \mathbf{x}_i - \boldsymbol{\mu}_j\rVert^2$

\State Assign sample: $c_i = \argmin_j d_j$

\EndFor

\For{$j=1$ {\bfseries to} $K$}

\State Update centroid:

$\boldsymbol{\mu}j = \frac{1}{n_j} \sum_{i=1}^{n} \mathbf{x}i \cdot \delta{c_i, j}$

\EndFor
\Until{convergence}
\end{algorithmic}
\end{algorithm}

\begin{figure}
    \centering
    \includegraphics[width=\textwidth]{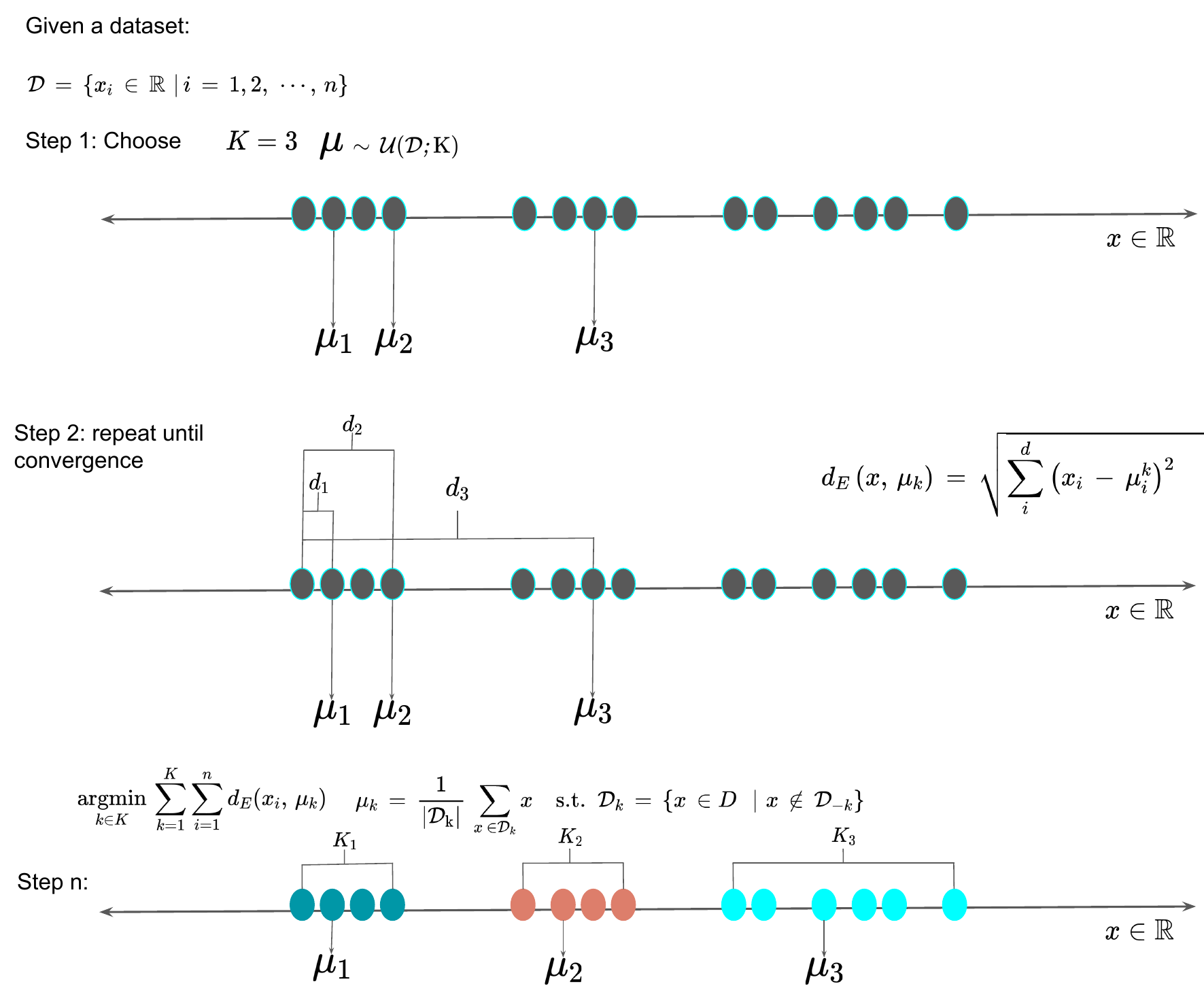}
    \caption{A visual illustration of $K$-Means clustering on a $1D$ dataset where we have $n=14$ samples and the cluster centriods, $K=3$. In this formulation, the initial centroids are uniformly randomly chosen as indicated by the vector means $\large{\bm{\mu}} \sim  \mathcal{U}(\mathcal{D}, K)$ as indicated in step 1. The distance $d_E(x_i, \large{\bm{\mu}}_i)$ is computed and the sample $x_i$ is assigned to the cluster with the minimum distance as shown in the figure where sample $x_1$ is assigned to cluster $1$. The new cluster centroids as computed from the updated cluster assignment such that the mean $\large{\bm{\mu}}_k$ of cluster $k$ is all samples that belong to that cluster. This step is repeated until convergence. Finally, all samples have been properly clustered as indicated by step n in the figure where the colors indicate the clusters to which each data point belongs.}
    \label{fig:K_Means_Clustering}
\end{figure}
The algorithm iterates until convergence, typically determined by a stopping criterion such as a maximum number of iterations or when the change in the centroids falls below a certain threshold. We have illustrated this approach to clustering a $1D$ feature space in Fig.~\ref{fig:K_Means_Clustering}.
K-means clustering is a computationally efficient algorithm and can handle large datasets. However, its performance can be sensitive to the initial choice of centroids and the number of clusters. Additionally, k-means clustering assumes that clusters are spherical and equally sized, which may not always be true in practice.
\subsection{Weighted Patch Matching}
Weighted Patch Matching (WPM) is an effective technique for measuring similarity which can be applied to medical image analysis~\cite{guo2015deformable,cordier2015patch}. It is based on matching small image patches between a reference image $I_R$ and a target image $I_T$ using weighted feature vectors~\cite{mechrez2016patch}. To do this, we first extract corresponding patches $P_i^R$ and $P_j^T$ from the reference and target images respectively. Each patch is represented by a feature vector $f_i^R$ and $f_j^T$ containing intensity, gradient, or other information. To provide robustness against intensity variations, a weight vector $w_i^R$ is computed for each reference patch based on the gradient magnitude:
\[
w_i^R(k) = \frac{||\nabla I_R(i+k)||_2}{\max\limits_k ||\nabla I_R(i+k)||_2},
\]
where $k$ indexes the pixels in the patch $P_i^R$. The weighted feature vector is:
\[
\hat{f}_i^R = w_i^R \odot f_i^R.
\]
The distance between two patches is then computed as:  
\[ 
d(P_i^R, P_j^T) = ||\hat{f}_i^R - f_j^T||_2.
\]
By minimizing $d$, we find corresponding patches that match well in terms of weighted features, even if the raw intensity values differ. We have shown the conceptual steps for WPM in Algorithm~\ref{alg:WPM}.

WPM is well-suited for medical image analysis tasks like registration and segmentation. It can match patches despite differences in scanner modalities or image intensities. The gradient-based weights emphasize edges and textures while suppressing flat regions. This focuses the matching on perceptually relevant structures. For example, WPM has been applied to align magnetic resonance (MR) and transrectal ultrasound (TRUS) images for targeted prostate biopsy. It outperformed standard patch matching due to its ability to handle intensity variations between modalities~\cite{wu2012brain}. WPM has also been used to improve atlas-based segmentation of abdomen CT scans, providing accuracy comparable to interactive segmentation~\cite{tong2015discriminative}. WPM often provides more flexibility in selecting similar patches by assigning different weights based on their relevance or similarity. This can improve the accuracy of patch matching, especially in images with varying noise levels or complex textures. However, WPM can be computationally intensive due to the calculation of weights for each patch. Additionally, the method's effectiveness can heavily depend on the choice of the weighting function, which may require careful tuning for different types of images.

\begin{algorithm}
\caption{Weighted Patch Matching (WPM)}
\label{alg:WPM}

\begin{algorithmic}

\State Extract patches $P_i^R$ from reference image $I_R$ and $P_j^T$ from target image $I_T$

\For{$i = 1$ to $N_R$}

\State Compute feature vector $f_i^R$ for patch $P_i^R$

\State Compute weight vector $w_i^R$ based on $\nabla P_i^R$

\State Compute weighted feature $\hat{f}_i^R = w_i^R \odot f_i^R$

\For{$j = 1$ to $N_T$}

\State Compute distance $d(P_i^R,P_j^T)$ between $\hat{f}_i^R$ and $f_j^T$

\EndFor

\State Find $k = \argmin_j d(P_i^R,P_j^T)$

\State Match $P_i^R$ to $P_k^T$

\EndFor

\end{algorithmic}
\end{algorithm}

Overall, WPM is a versatile technique for medical image-matching problems where robustness to intensity variations is critical. Its continued development, including learning-based weight optimization and integration with deep networks, will likely extend its applicability even further.

\subsection{Block Matching}
Block Matching is a widely used technique for estimating correspondences between images by matching small local blocks or patches. In medical imaging, it can be highly effective for tasks like inter-frame motion tracking and deformable image registration. One particularly useful variant utilizes the $l_1$ norm as the block similarity metric to provide robustness against outliers~\cite{sotiras2011deformable}.

The key idea in Block Matching is to divide the reference image $I_R$ and target image $I_T$ into smaller blocks $B_i^R$ and $B_j^T$ of pixels respectively, as shown in Algorithm~\ref{alg:bm} lines 1-2. The goal is then to find correspondences between individual blocks from $I_R$ and semantically similar blocks in $I_T$. To enable numeric comparison, each image block is represented by a feature vector $f_i^R$ and $f_j^T$ capturing intensities, gradients, textures, or other descriptive features (see lines 4-9, Algorithm \ref{alg:bm}).

A distance metric between feature vectors is then used to quantify block similarity. Using the $l_1$ norm, $\lVert\cdot\lVert_1$ defined as:
\begin{flalign}
    d(B_i^R, B_j^T) &= \lVert f_i^R - f_j^T\rVert_1 \\
    &= \sum_{k=1}^{n} \mid f_i^R(k) - f_j^T(k)\mid ,
\end{flalign}
provides a robust block comparison that is less influenced by outlier feature differences compared to the $l_2$ norm~\cite{sotiras2011deformable}. Here, $n$ denotes the dimensionality of the block feature vectors. By minimizing the $l_1$ distance (lines 12-19), the best matching target block $B_j^T$ can be found for each reference block $B_i^R$~\cite{ledesma2005motion}.

Applying Block Matching with $l_1$ similarity to medical images is advantageous as intensity values can vary considerably across modalities and between scans due to noise, artifacts, and pathologies. The $l_1$ metric provides a degree of invariance to such complex conditions.

In summary, Block Matching using $l_1$ block feature distance is a versatile and robust technique for medical motion estimation and correspondence problems~\cite{sotiras2011deformable,ledesma2005motion}. Block matching is often simpler and faster, making it suitable for real-time applications or scenarios with limited computational resources. It is effective in cases where the similarity between patches is relatively uniform. However, it might lack the adaptability of WPM when dealing with complex or non-homogeneous noise patterns. The fixed criteria for selecting similar patches may lead to suboptimal matching in images with high variability, potentially resulting in less effective denoising or reconstruction.

\begin{algorithm}
\caption{Block Matching with $l_1$ norm}
\label{alg:bm}
\begin{algorithmic}[1]

\State Divide reference image $I_R$ into blocks $B_i^R$
\State Divide target image $I_T$ into blocks $B_j^T$

\For{$i = 1$ to $N_R$}
\State Compute feature vector $f_i^R$ for block $B_i^R$
\EndFor

\For{$j = 1$ to $N_T$}
\State Compute feature vector $f_j^T$ for block $B_j^T$
\EndFor

\For{$i = 1$ to $N_R$}
\State Initialize best match distance $d_{\min} = \infty$
\For{$j = 1$ to $N_T$}

\State Compute $l_1$ distance $d(B_i^R, B_j^T) = ||f_i^R - f_j^T||1$
\If{$d(B_i^R, B_j^T) < d_{\min}$}
\State $d_{\min} = d(B_i^R, B_j^T)$ 
\State $match(B_i^R) = B_j^T$
\EndIf
\EndFor
\EndFor

\end{algorithmic}
\end{algorithm}

\section{Low-rank methods}{\label{sec:low-rank}}
This section delves into the mathematics and applications of the low-rank techniques within the studies reviewed in Sections~\ref{sec:LRMA} and \ref{sect:LLRMA}. The reviewed papers used several types of low-rank techniques such as SVD, weighted SVD, randomized SVD, tensor SVD, and PCA across various medical image data modalities to facilitate tasks such as feature extraction, compression, denoising, decomposition, and more. The low-rank method provides a solid framework for analyzing and processing high-dimensional data by extracting the most relevant information while reducing computational complexity and/or noise.

\subsection{Traditional SVD}
\begin{figure}
    \centering
    \includegraphics[scale=0.45]{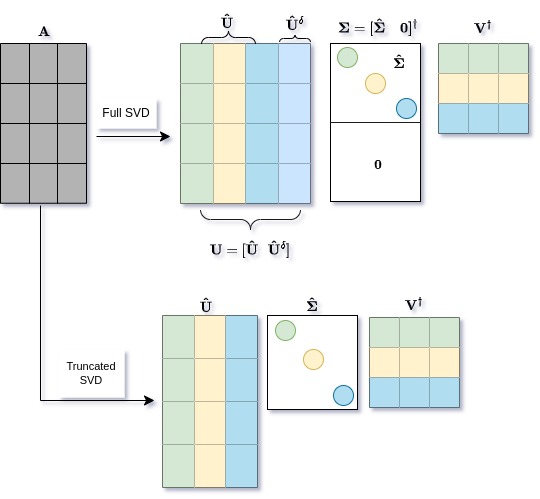}
    \caption{A demonstration of full and truncated SVD. The columns of $\hat{\mathbf{U}}^\delta$ span a vector space that is complementary and orthogonal to that spanned by $\hat{\mathbf{U}}$ and $\hat{\mathbf{\Sigma}}$ is an $m\times m$ diagonal matrix.}
    \label{fg:SVD}
\end{figure}

{\label{LRA}}  
When an image can be approximated sufficiently well by a low-rank matrix, SVD can be used to find this approximation. This LORMA can be represented more compactly than the original image. SVD decomposes a complex data matrix $\mathbf{A}\in \mathbb{C}^{m\times n}$ into three independent matrices \cite{wall2003singular, stewart1993early, klema1980singular}:
\begin{flalign}{\label{SVD1}}
    \mathbf{A}&=\mathbf{U} \mathbf{\Sigma} \mathbf{V}^{\dagger}\\
    &=\left[\hat{\mathbf{U}}\,\,\,\hat{\mathbf{U}}^\delta\right] \begin{bmatrix}
\hat{\mathbf{\Sigma}} & \mathbf{0}\\
\mathbf{0} & \mathbf{0}
\end{bmatrix}\mathbf{V}^{\dagger},
\end{flalign}
where $\mathbf{U}\in\mathbb{C}^{m\times m}$ and $\mathbf{V}\in\mathbb{C}^{n \times n}$ are unitary matrices, $\mathbf{\Sigma}=\text{diag}(\xi_1, \xi_2,\cdots, \xi_{m})\in \mathbb{C}^{m \times n}$ is a diagonal matrix of rank $m$ and $^\dagger$ denotes a conjugate transpose. The singular values of the matrix $\mathbf{A}$, which are denoted as $\xi_i$ and lie on the diagonal entries of the matrix $\mathbf{\Sigma}$, are uniquely determined by $\mathbf{A}$. These singular values are arranged in a descending order, that is $\xi_1\geq \xi_2\geq \xi_3\cdots \geq\xi_{m}\geq 0$. It is easy to show that $\lambda_i=\sqrt{\xi_i}$ where $\lambda_i$ denotes the eigenvalues of $\mathbf{AA}^\dagger$ or $\mathbf{A}^{\dagger}\mathbf{A}$. The columns of $\mathbf{U}$ and the columns of $\mathbf{V}$ are the left and right singular vectors of $\mathbf{A}$, respectively. The vectors $\mathbf{u}_i \in \mathbb{C}^{m}$ and $\mathbf{v}_i\in \mathbb{C}^{n}$ are the orthonormal eigenvectors of $\mathbf{AA}^{\dagger}$ and  $\mathbf{A}^{\dagger}\mathbf{A}$ respectively. Therefore, 
Eq.~\ref{SVD1} can be written as follows 
\begin{gather}
 \mathbf{A}=\sum_{i=1}^{m}\xi_i\mathbf{u}_i\mathbf{v}_i^{\dagger}.
\end{gather}
If there exists a matrix of rank $r$ such that $0\leq r \leq m$, then the optimal approximation of $\mathbf{A}$ that preserves the most important features is given by
\begin{gather}
 \mathbf{A}_r=\sum_{i=1}^{r}\xi_i\mathbf{u}_i\mathbf{v}_i^{\dagger}.
\end{gather}
When $r=m$ then $\mathbf{A}=\mathbf{A}_r$. Each value of $\xi_i$ quantifies how much the corresponding component $\xi_i \mathbf{u}_i\mathbf{v}_i^{\dagger}$ contributes in the reconstruction of $\mathbf{A}$, that is, the larger the value of $\xi_i$ the more $\xi_i \mathbf{u}_i\mathbf{v}_i^{\dagger}$ contributes to the reconstruction of $\mathbf{A}$. The SVD technique constructs a LORMA matrix $\mathbf{A}_r$ by solving the following optimization problem:
\begin{gather}
\underset{\mathbf{A}_r,\,\,\text{s.t.}\,\, \text{rank}(\mathbf{A}_r)=r}{\mathrm{argmin}} \,\lVert \mathbf{A}- \mathbf{A}_r\rVert_F=\mathbf{U}_r\mathbf{\Sigma}_r \mathbf{V}_r^\dagger,
\end{gather}
where $\mathbf{U}_r, \mathbf{V}_r$ denotes the first $r$ leading columns of $\mathbf{U}$ and $\mathbf{V}$, $\mathbf{\Sigma}_r$ denotes a sub-matrix of $\mathbf{\Sigma}$ and $\lVert \cdot\rVert_F$ denotes the Frobenius norm. Fig.~\ref{fg:SVD} demonstrates the steps in full and truncated SVD. 

The main strength of SVD is its ability to reduce the dimensionality of data, which is particularly useful for tasks involving large datasets. By approximating complex matrices with simpler matrices, SVD enables efficient feature extractions, compression and noise reduction, improving signal quality and/or image processing techniques. However, SVD has its drawbacks. The computational cost of full SVD is relatively high in terms of time and memory. Its time complexity is typically $\mathcal{O}(m n\min\{m, n\})$ \cite{holmes2007fast, mamat2007statistical, brand2006fast, roughgarden2015cs168, atemkeng2023lossy, halko2011finding}. In terms of memory, it requires $\mathcal{O}(mn)$ to store the three component matrices \cite{roughgarden2015cs168, burca2014fast, atemkeng2023lossy, halko2011finding}. The full SVD provides a complete representation of the original matrix and preserves all information that may be important for certain applications. However, it can be impractical for large datasets due to the high computational requirements. The truncated SVD, on the other hand, approximates the original matrix by retaining only $r$ singular values and their corresponding columns from $\mathbf{U}_r$ and $\mathbf{V}^\dagger_r$. This significantly reduces the computational cost. The time complexity of truncated SVD is approximate $\mathcal{O}(mnr)$ \cite{brand2006fast, roughgarden2015cs168, burca2014fast, atemkeng2023lossy, halko2011finding}, and its memory complexity is $\mathcal{O}(r(m+ n))$ \cite{aishwarya2016lossy, roughgarden2015cs168, burca2014fast, atemkeng2023lossy}. The truncated SVD is well suited for dimensionality reduction and noise reduction tasks, where the lower-dimensional approximation captures the most important features of the data. However, some information is discarded in the process, which can lead to a loss of accuracy in certain applications. In addition, the assumption of orthogonal data in SVD may not be true for all datasets, limiting its effectiveness in capturing the underlying structures.
\subsection{Weighted SVD}
\begin{figure}
    \centering
    \includegraphics[scale=0.45]{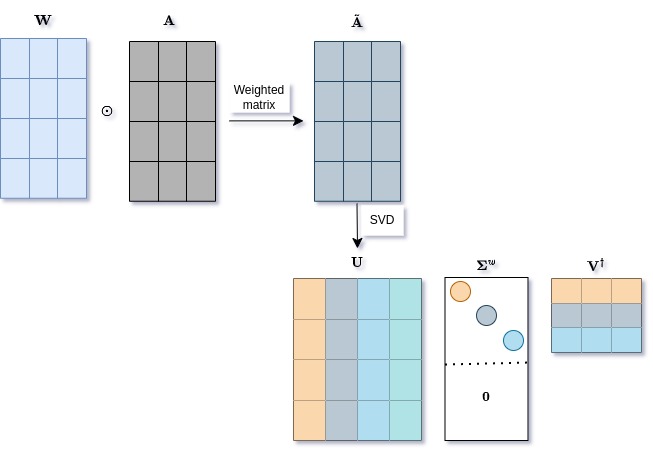}
    \caption{A demonstration of weighted SVD.}
    \label{fg:wSVD}
\end{figure}

Weighted SVD is a method for matrix factorization that extends the standard SVD method \cite{van1976generalizing}. The weighted matrix, $\tilde{\mathbf{A}}$ with respect to the diagonal weight matrix $\mathbf{W}$ is defined as
\begin{align}
\tilde{\mathbf{A}}=\mathbf{W}\odot \mathbf{A},
\end{align}
where $\odot$ denotes the Schur product \cite{paulsen1989schur}. The weighted SVD of $\tilde{\mathbf{A}}$ is expressed as
\begin{align}
    \tilde{\mathbf{A}}=\mathbf{U}\mathbf{\Sigma}^w\mathbf{V}^{\dagger},
\end{align}
where $\mathbf{\Sigma}^w$ denotes the diagonal matrix of weighted singular values, which is obtained by scaling the diagonal matrix $\mathbf{\Sigma}$ with the weights
\begin{align}
 \mathbf{\Sigma}^w_{jj}=\mathbf{W}_{jj}\mathbf{\Sigma}_{jj}.
\end{align}
 The decomposition solves the optimization problem:
\begin{gather}
\underset{\tilde{\mathbf{A}},\,\,\text{s.t.}\,\, \text{rank}(\tilde{\mathbf{A}})=r}{\mathrm{argmin}} \,\lVert \mathbf{A}- \tilde{\mathbf{A}}\rVert_F=\mathbf{U}\mathbf{\Sigma}^w\mathbf{V}^{\dagger}
\end{gather}
Fig.~\ref{fg:wSVD} illustrates how the weighted SVD works. The weighted SVD technique often involves applying weights to a matrix, which modifies the importance of certain rows or columns. This impacts how singular values are adjusted during decomposition. In image compression, for example, weights are assigned based on the importance of each pixel in the image. In other words, pixels with higher contrast may be given more weight than pixels with lower contrast.

Weighted SVD  accurately captures the underlying structure of the data by weighting singular values according to their importance. This can be particularly beneficial in the case of noisy or incomplete data, as the algorithm can give more weight to reliable information. In addition, weighted SVD allows for better interpretability and control over the decomposition process, as particular singular values can be assigned a higher or lower priority based on domain knowledge. However, the computational cost of weighted SVD is generally higher than that of the traditional SVD due to the added complexity of handling weights. The weighted SVD introduces additional operations to account for the weights, which can lead to a higher time complexity depending on the specific implementation \cite{chen2017weighted}. The memory requirement is the same as for truncated or full SVD since $\mathbf{W}_{jj}\mathbf{\Sigma}_{jj}$ can be stored directly.
\subsection{Randomized SVD}
Randomized Singular Value Decomposition (rSVD) refers to a technique that approximates the SVD of a matrix using random projections \cite{halko2011finding}. The rSVD method has gained increasing popularity in recent years owing to its efficiency and scalability in large-scale data analysis \cite{mahoney2011randomized, halko2011finding, gower2015randomized}. In a general LORMA problem, the objective is to approximate the provided matrix $\mathbf{A}$ using two smaller matrices, $\mathbf{B}\in \mathbb{C}^{m\times r}$ and $\mathbf{C}\in \mathbb{C}^{r\times n}$ with the conditions $r\ll n$ and $\mathbf{A}\approx \mathbf{B}\mathbf{C}$. To achieve this approximation through rSVD, a two-stage computation was proposed by \cite{halko2011finding}. In the initial stage, a random projection matrix $\mathbf{P}\in\mathbb{C}^{n\times r}$ (with entries typically drawn from a standard normal or uniform distribution) is created to sample the column space of $\mathbf{A}\in \mathbb{C}^{m\times n}$, represented as:
\begin{align}
    \mathbf{Z}=\mathbf{A}\mathbf{P},
\end{align}
where $\mathbf{Z}\in \mathbb{C}^{m\times r}$ may be smaller than $\mathbf{A}$, especially for low-rank matrices with $r\ll n$. As a random projection matrix $\mathbf{P}$ is unlikely to eliminate crucial components of $\mathbf{A}$, $\mathbf{Z}$ serves as an accurate approximation of the column space of $\mathbf{A}$ \cite{brunton2022data}. Consequently, the low-rank QR decomposition of $\mathbf{Z}$ can be computed to obtain an orthonormal basis for $\mathbf{A}$:
\begin{align}
    \mathbf{Z}=\mathbf{Q}\mathbf{R}.
\end{align}
In the second step, using the low-rank basis $\mathbf{Q}$, $\mathbf{X}$ is projected into a smaller space, expressed as:
\begin{align}
    \mathbf{Y}=\mathbf{Q}^\dagger\mathbf{A}.
\end{align}
The SVD of $\mathbf{Y}$ is given by:
\begin{align}
    \mathbf{Y}=\hat{\mathbf{U}}\mathbf{\Sigma}\mathbf{V}^\dagger.
\end{align}
As $\mathbf{Q}$ is orthonormal and approximates the column space of $\mathbf{A}$, the matrices $\mathbf{\Sigma}$ and $\mathbf{V}$ are the same for both $\mathbf{Y}$ and $\mathbf{A}$. The high-dimensional left singular vectors $\mathbf{U}$ are reconstructed using $\hat{\mathbf{U}}$ and $\mathbf{Q}$:
\begin{align}
    \mathbf{U}=\mathbf{Q}\hat{\mathbf{U}}.
\end{align}
\begin{figure}
    \centering
    \includegraphics[scale=0.42]{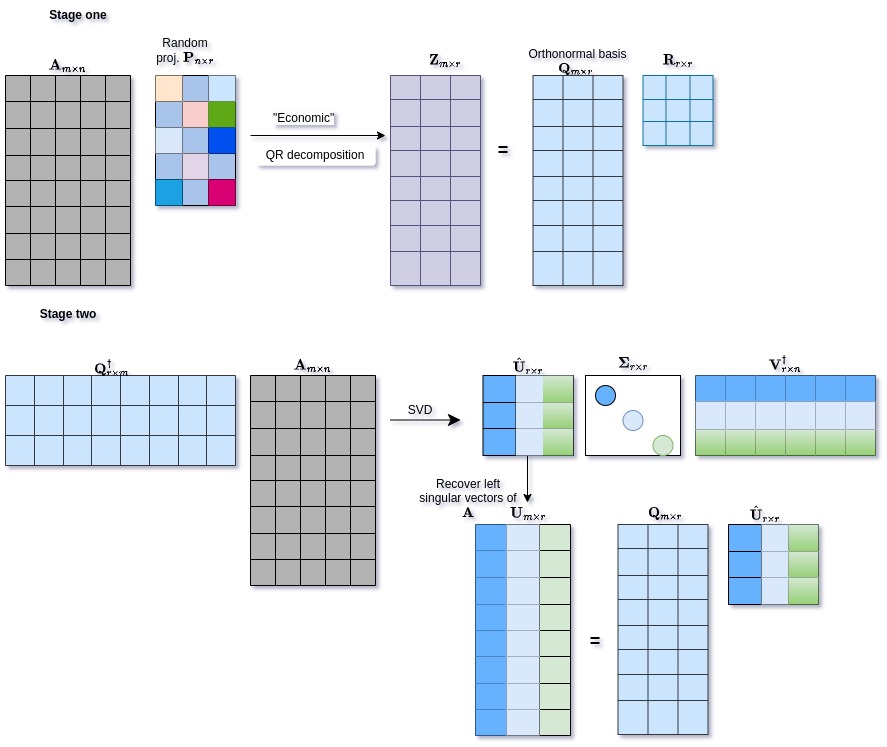}
    \caption{A demonstration of random SVD.}
    \label{fig:rSVD}
\end{figure}
The rSVD algorithm is based on the idea that the matrix $\mathbf{Q}$ acts as a randomized projection matrix, compressing the information in $\mathbf{A}$ to a smaller dimension, which is then used to compute the SVD of $\mathbf{A}$. The accuracy of the approximation depends on the number of columns, denoted as $r$, in the matrix $\mathbf{Q}$. A larger value of $r$ leads to increased accuracy but also comes with higher computational cost. Fig.~\ref{fig:rSVD} demonstrates the steps in rSVD.

rSVD offers a computational advantage over traditional SVD by significantly reducing the time complexity associated with matrix factorization, especially for large datasets. The computational cost of rSVD is approximately given by $\mathcal{O}(m n \log(r))$ \cite{halko2011finding}. The logarithmic term ($\log(r)$) represents the number of iterations required for the power iteration method to converge to the desired number of singular values or components. In addition, rSVD has lower memory requirements, with a memory complexity of about $\mathcal{O}(mn + nr)$ \cite{singh2020range, halko2011finding}. These attributes make rSVD highly scalable and efficient for applications with large amounts of data \cite{mahoney2011randomized, halko2011finding, gower2015randomized}. Moreover, due to the introduction of randomness, rSVD can provide higher robustness to outliers and noise, leading to more stable results in some instances. However, rSVD comes with certain limitations. It provides an approximate decomposition that may not achieve the same level of accuracy as the exact decomposition provided by traditional SVD. The need for parameter tuning, including selecting appropriate values for parameters such as the number of samples and the desired rank, may affect the approximation quality and must be carefully considered. It should also be noted that rSVD is best suited for tasks where an approximate decomposition is acceptable, such as dimensionality reduction. It may not be suitable for applications that require exact results.

\subsection{Tensor SVD}
The tensor SVD works by unfolding the tensor along one of its modes (i.e., dimensions), resulting in a tensor matrix representation \cite{kilmer2021tensor}. The matrix SVD is then applied to this matrix representation to produce a set of orthogonal matrices and singular values. Let $\mathcal{A}\in \mathbb{C}^{I_1\times I_2\times\cdots\times I_n}$ denotes a tensor of order $n$. The decomposition of a tensor, $\mathcal{A}$, is given by
\begin{align}{\label{tensor}}
    \mathcal{A} = \mathcal{B} \times_1 \mathbf{U}^{(1)} \times_2 \mathbf{U}^{(2)} \times_3 \cdots \times_n \mathbf{U}^{(n)},
\end{align}
where $\mathcal{B}$ is a core tensor of size $(r_1, r_2, \cdots r_n)$ with rank $r$, and $\mathbf{U}^{(1)}, \mathbf{U}^{(2)}, \cdots, \mathbf{U}^{(n)}$ are orthogonal matrices with size $I_1 \times r_1, I_2 \times r_2, \cdots, I_n \times r_n$ respectively. The weights associated with each tensor component are stored in the core tensor $\mathcal{B}$, and the orthogonal matrices $\mathbf{U}^{(1)}, \mathbf{U}^{(2)}, \cdots, \mathbf{U}^{(n)}$ represent the contribution of each mode to each component. The number of components is determined by $r$, which is typically much smaller than the tensor size. The components are recorded in descending order, with the first capturing the most variation in the tensor. 
Eq.~\ref{tensor} can be represented using a matrix form as
\begin{align}
\mathcal{A}= \sum_{i_1=1}^{r_1} \sum_{i_2=1}^{r_2} \cdots \sum_{i_n=1}^{r_n} b_{i_1,i_2,\cdots,i_n} \mathbf{u}_{i_1}^{(1)}\otimes \mathbf{u}_{i_2}^{(2)} \cdots\otimes \mathbf{u}_{i_n}^{(n)},    
\end{align}
where $\mathbf{u}_{i_j}^{(n)}$ denotes an $i^{th}$ column of $\mathbf{U}^{(n)}$ and $\otimes$ denotes the tensor product operation. If $b_{i_1,i_2,\cdots,i_n}=0$, that is, $\mathcal{B}$ is diagonal then
\begin{align}{\label{mtSVD}}
    \mathcal{A}= \sum_{i_1=1}^{r} b_{ii\cdots i} \mathbf{u}_{i}^{(1)}\otimes \mathbf{u}_{i}^{(2)} \cdots\otimes \mathbf{u}_{i}^{(n)}
\end{align}
except when $i_1=i_2=\cdots =i_n$. The t-SVD is defined as
\begin{align}
        \mathcal{A}= \sum_{i_1=1}^{r} \xi_i \mathbf{u}_{i}^{(1)}\otimes \mathbf{u}_{i}^{(2)} \cdots\otimes \mathbf{u}_{i}^{(n)},
\end{align}
where $\xi_i$ is the $i^{th}$ singular value. Eq.~\ref{mtSVD} underlines the similarity between the t-SVD and the matrix SVD, where the singular values are replaced by tensors, and the factor matrices are replaced by vectors.

The t-SVD technique has both advantages and disadvantages in various applications. One of its notable advantages is its ability to reduce the dimensionality of multidimensional data, which is useful in compression, noise reduction, and feature extraction. This technique proves valuable in data analysis, as it helps uncover intricate patterns and relationships in complex datasets. However, tensor SVD also has drawbacks. As with any dimensionality reduction method, t-SVD can result in information loss since only the most significant singular values and vectors are preserved. While it can reveal patterns, their interpretation can be difficult, especially for high-dimensional tensors. In addition, the computational cost can be significant, especially for high-dimensional tensors, making it less practical for large datasets. The computational complexity of t-SVD in some typical tensor decomposing-based methods is about $\mathcal{O}(n^2m^2k+k^3)$, where $n$, $m$, and $k$ denote the tensor dimensions \cite{wang2023tensor}.
\subsection{Principal Component Analysis}
\begin{figure}
    \centering
    \includegraphics[scale=0.45]{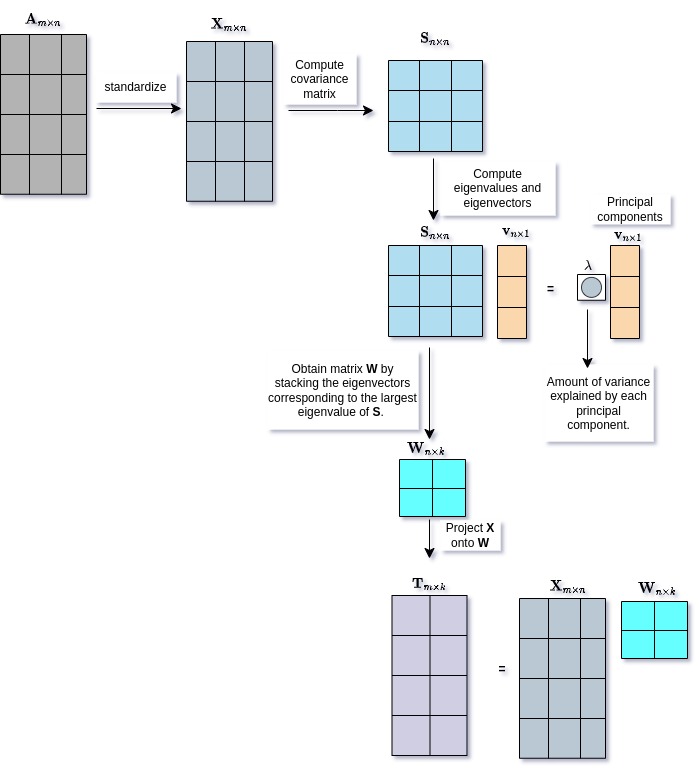}
    \caption{A demonstration of PCA In this example, $k$ denotes the number of principal components to retain.}
    \label{fig:pca}
\end{figure}

Principal component analysis (PCA) is a mathematical technique used in data analysis and ML for dimensionality reduction and feature extraction. It identifies the most important features, or principal components, in the data matrix $\mathbf{A}$ that capture the most significant variance. When reducing the number of feature variables in a dataset, reduced dimensionality trades accuracy for simplicity. Small datasets are easier to explore, visualize, and analyze with ML algorithms since they have few variables. Let $\mathbf{x}_i$ denote the $i^{th}$ column vector of the data matrix $\mathbf{X}$ (denotes a covariance matrix of the centered variable) with $n$ feature variables. A linear combination of $\mathbf{x}$'s is expressed as
\begin{align}
    \mathbf{t}&=\sum_{i=1}^{n}w_i\mathbf{x}_i\\
    &=\mathbf{X}\mathbf{w}.
\end{align}
The objective is to find a set of orthogonal linear combinations of variables that capture the most variation in the data. That is,
\begin{align}
    \underset{\lVert\mathbf{w}\rVert=1}{\operatorname{argmax}}\,\, \text{var}(\mathbf{\mathbf{X}\mathbf{w}}).
\end{align}
To compute the first principal component, the first weight vector, $\mathbf{w}_{(1)}$, has to satisfy
\begin{align}{\label{maxvar}}
  \mathbf{w}_{(1)}&=  \underset{\lVert\mathbf{w}\rVert=1}{\operatorname{argmax}}\,\, \lVert\mathbf{X}\mathbf{w}\rVert^2\\
  &=\underset{\lVert\mathbf{w}\rVert=1}{\operatorname{argmax}}\,\, \left(\mathbf{\mathbf{w}^\dagger\mathbf{X}^\dagger \mathbf{X}\mathbf{w}}\right).
\end{align}
Since $\mathbf{w}$ is a unit vector, Eq.~\ref{maxvar} can be rewritten as
\begin{align}
    \mathbf{w}_{(1)}=\underset{}{\operatorname{argmax}}\,\, \left(\frac{\mathbf{\mathbf{w}^\dagger\mathbf{X}^\dagger \mathbf{X}\mathbf{w}}}{\mathbf{w}^\dagger\mathbf{w}}\right).
\end{align}
To compute the $k^{th}$ component, the $k-1$ principal components are subtracted from $\mathbf{X}$, that is
\begin{align}
    \tilde{\mathbf{X}}_k=\mathbf{X}-\sum_{j=1}^{k-1}\mathbf{X}\mathbf{w}_{(j)}\mathbf{w}_{(j)}^\dagger
\end{align}
and the weight vector that extracts maximum variance from $\tilde{\mathbf{X}}_k$ is given by
\begin{align}
     \mathbf{w}_{(k)}&=  \underset{\lVert\mathbf{w}\rVert=1}{\operatorname{argmax}}\,\, \lVert\tilde{\mathbf{X}}_k\mathbf{w}\rVert^2\\
     &=\underset{\lVert\mathbf{w}\rVert=1}{\operatorname{argmax}}\,\, \left(\mathbf{w}^\dagger\tilde{\mathbf{X}}^\dagger_k \tilde{\mathbf{X}}_k\mathbf{w}\right).
\end{align}
Let $\mathbf{W}$ denote a matrix containing the $k$ selected eigenvectors as its columns, then the full principal component decomposition can be computed as 
\begin{align}
    \mathbf{T}=\mathbf{X}\mathbf{W},
\end{align}
where $\mathbf{T}$ is the matrix of principal components and each column represents a principal component. The resulting matrix $\mathbf{T}$ contains the data's reduced-dimensional representation, capturing the most significant variation. The covariance matrix's eigenvectors and eigenvalues are critical in determining the directions and magnitudes of the principal components, respectively. Fig.~\ref{fig:pca} demonstrates the steps in PCA.

The main advantages of PCA include its ability to reduce the dimensionality of data while preserving important information, noise reduction, improve model performance, and visualize high-dimensional data. It is an unsupervised technique suitable for exploratory analysis. However, PCA assumes a linear data distribution and may not perform well with nonlinear relationships. It is also sensitive to outliers, as these can disproportionately affect the calculated principal components. In addition, interpretation of the transformed components can be challenging for complex datasets. The computation time for PCA is mainly influenced by the number of data points, $m$, and the number of features, $n$, in the dataset. The most time-consuming step is usually the eigenvalue decomposition or SVD, which has a computational complexity of $\mathcal{O}(n^3)$ for the covariance matrix and $\mathcal{O}(\min\{m^2n, mn ^2\})$ for SVD \cite{van2009dimensionality,li2019tutorial}. The memory requirement depends on the size of the covariance matrix or the SVD factor matrices. It is typically $\mathcal{O}(n^2)$ for the covariance matrix and $\mathcal{O}(\min\{m^2, mn\})$ for SVD \cite{van2009dimensionality,li2019tutorial}.

Table~\ref{tab:lowrank} provides a summary of various low-rank methods used in data analysis. It highlights their key strengths, weaknesses, and computational costs. The table is a valuable reference for selecting the most suitable low-rank method based on specific data analysis requirements.
\begin{landscape}
\pagestyle{empty}
\begin{table} 
  \begin{center}
    \caption{This table provides a detailed analysis of various low-rank methods, highlighting their respective strengths, weaknesses, computational cost, performance and recommended use.}
    \begin{tabular}{p{1.5cm} p{4.5cm} p{2.5cm} p{4.3cm} p{4cm}p{2cm}} 
        \hline
      \textbf{low-rank method} & \textbf{Strengths} & \textbf{Weakness}&\textbf{Computational cost}&\textbf{Performance}&\textbf{Recommended Use}\\
      \hline
      SVD&$\bullet$Provides an exact decomposition of the data matrix into orthogonal factors.\newline $\bullet$Useful for dimensionality reduction, noise reduction, and feature extraction.&
$\bullet$Sensitive to missing data and outliers.\newline $\bullet$May not handle high-dimensional data well.
  &$\bullet$ 
Full SVD has time complexity of $\mathcal{O}(m n\min\{m, n\})$ and  memory complexity of $\mathcal{O}(mn)$. While truncated SVD has time complexity $\mathcal{O}(mnr)$, and its memory complexity is $\mathcal{O}(mr + nr)$.&$\bullet$High accuracy in noise reduction and data compression, but less suitable for real-time applications due to high computational demand.&$\bullet$MRI, CT scans (global structure preservation)\\
      \hline
      weighted SVD&$\bullet$Accurately captures the underlying structure of the data by weighting singular values according to their importance.\newline$\bullet$Provides better interpretability and control over the decomposition process.&$\bullet$Computational cost is generally higher than that of the traditional SVD. &$\bullet$When $\mathbf{W}_{jj}\mathbf{\Sigma}_{jj}$ is stored directly, memory requirement is approximately $\mathcal{O}(mnr)$ or $\mathcal{O}(m n\min\{m, n\})$.&$\bullet$Improved interpretability and precision in feature extraction, but the added complexity can slow down performance in large-scale datasets.&$\bullet$Ultrasound, low-dose CT (detail preservation)\\
      \hline
      rSVD&$\bullet$Faster than computing the full SVD for large matrices.\newline $\bullet$Extract the most important features of a matrix while discarding noise and less significant components. &$\bullet$SVD might not have clear interpretations as they would with the traditional SVD.&$\bullet$ Has time complexity of approximately $\mathcal{O}(m n \log(r))$ and a memory complexity of about $\mathcal{O}(mn + nr)$.&$\bullet$Good speed and efficiency for large datasets, suitable for real-time processing where precision is less critical.&$\bullet$Large-scale MRI, CT (fast denoising)\\
      \hline
      t-SVD&$\bullet$ Reduces the dimensionality of multidimensional data, which is useful in compression, noise reduction, and feature extraction &$\bullet$Can result in information loss since only the most significant singular values and vectors are preserved.&$\bullet$Complexity for full tensor SVD can be approximated as $\mathcal{O}(n^2m^2k+k^3)$.&$\bullet$Strong performance in handling tensor data, effective in compressing 3D and 4D data, but may struggle with very large data volumes.&$\bullet$Dynamic MRI, PET-CT (multi-dimensional data)\\
      \hline
      PCA&$\bullet$Has the ability to reduce the dimensionality of data while preserving important information, reduce noise reduction, improve model performance, and visualize high-dimensional data. &$\bullet$Sensitive to scaling of features.\newline$\bullet$It is not well suited for capturing nonlinear relationships in data.&$\bullet$Has a computational complexity of $\mathcal{O}(n^3)$ for the covariance matrix and $\mathcal{O}(\min\{m^2n, mn ^2\})$ for SVD. Memory complexity is typically $\mathcal{O}(n^2)$ for the covariance matrix and $\mathcal{O}(\min\{m^2, mn\})$ for SVD.&$\bullet$Reliable for general-purpose data reduction and visualization, but less effective with complex nonlinear patterns.&$\bullet$MRI, CT (dimensionality reduction)\\
    \label{tab:lowrank}
     \end{tabular}
  \end{center}
\end{table}
\end{landscape}

\section{Findings and Discussion}
\begin{figure}
    \centering
    \includegraphics[scale=0.45]{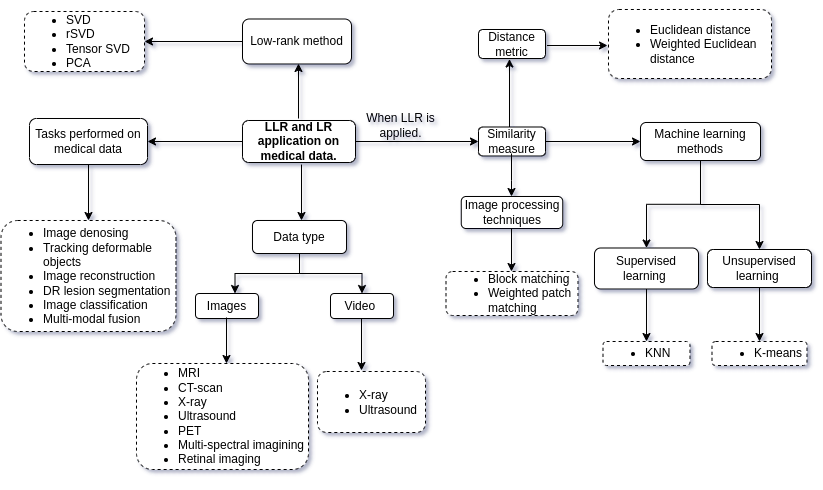}
    \caption{An overview of the application of LLORMA and LORMA to medical data.}
    \label{fg:SWork}
\end{figure}

{\label{sec:findings}}
This section addresses the findings and drawbacks that arise from the review. We also provide valuable insights and recommendations for possible future directions in this area of study. Fig.~\ref{fg:SWork} summarizes the findings of this study. The figure shows the low-rank method used in the reviewed publications while performing various tasks on the medical data. It also highlights the different types that were used. In terms of the LLORMA approximation, the figure shows the similarity measurement methods used in the studied publications during the matching phase.
\subsection{RQN1 and RQN2}
From the reviewed papers, LLORMA has shown the potential and ability to capture the underlying structure of data while reducing its dimensionality with little to no feature loss or distortion due to the locality coherence search. This approach balances data compression and detail preservation by exploiting the local low-rank property. The local nature of the approximation allows for localized analysis, where specific regions or patches of the data can be accurately represented, leading to improved performance in tasks such as image processing and computer vision \cite{mallat2016understanding}. Moreover, LLORMA exhibits robustness to outliers \cite{mozaffari2021robust}, which enhances its applicability in real-world scenarios where noisy or corrupted data may be present. This robustness ensures that the approximation remains accurate and reliable, even in the presence of abnormalities. Additionally, LLORMA provides a denoising capability, separating the essential signal from noise and improving the quality of the data representation \cite{xia2017denoising, lei2018denoising, zhai2018weighted, lv2019denoising, ji2020medical, chen2021novel}. Its effectiveness in preserving meaningful information while reducing computational complexity and memory usage makes LLORMA a powerful tool in various domains, including medical image analysis as studied in this paper. 

\begin{figure}
  \centering
  \includegraphics[scale=0.5]{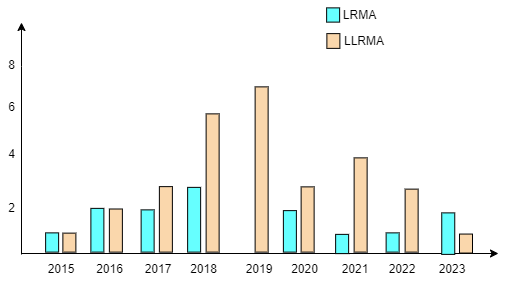}
  \caption{Number of publications that applied LORMA or LLORMA technique on medical images.}
  \label{fig:LLRMA/LRMA}
\end{figure}

\begin{figure}
\centering
  \includegraphics[scale=0.5]{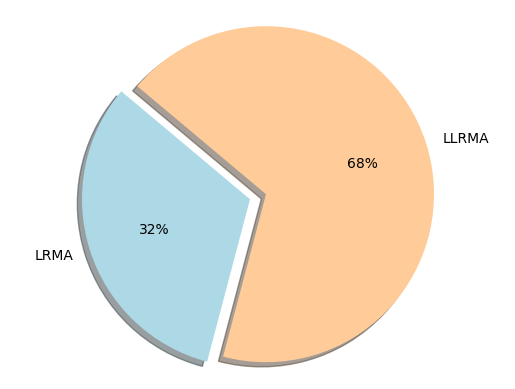}
  \caption{A pie chart showing the distribution of publications using LLORMA and LORMA .}
  \label{fig:pie}
  \end{figure}

It is important to note that LLORMA was originally introduced in 2013, but its use in the medical field began in 2015. This information is key to understanding the use of the LLORMA approximation in the medical field. Fig.~\ref{fig:LLRMA/LRMA} shows that there is no documented use of LLORMA prior to 2015. However, there was a significant change after 2015, when researchers in the medical field began to apply the LLORMA technique. During this period, we found that of the total 44 papers published in the medical field between 2015 and 2023, only 14 papers applied LORMA, while 30 papers used LLORMA techniques. Fig.~\ref{fig:pie} also shows that a more significant proportion of publications from 2015 to 2023 focused on the LLORMA approximation compared to the LORMA approximation, i.e., 32\% of publications used the LORMA approximation, while 68\% used the LLORMA approximation. This trend highlights a notable shift in research preferences and the crucial role that the LLORMA approximation plays in the medical field.
\subsection{RQN3: Application of LLORMA to Various Medical Imaging Modalities}
When applying LLORMA to different medical imaging modalities, the differences in imaging techniques, tissue contrasts, and noise characteristics play a critical role in determining the patch size and effective denoising technique. Each modality presents unique challenges due to the nature of the data and noise, requiring customized approaches. For MRI, medium to large patches are often preferred due to the complex structures and varying tissue contrasts in these images. Larger patches can better capture these variations, while medium-sized patches help to avoid boundary artifacts. Studies such as \cite{fu20163d} (tensor approximations) and \cite{xia2017denoising} (WNNM for Rician noise) support this approach and show that techniques such as non-local means and tensor SVD with total variation effectively preserve structural details. CT imaging, on the other hand, benefits from smaller patches to preserve fine details and edges that are crucial for accurate diagnosis in high-resolution images. Managing Poisson noise, especially at low doses, is a major concern. Techniques such as t-SVD and sparse coding with low-rank constraints, as demonstrated in \cite{sagheer2019denoising} and \cite{lei2018denoising}, help to balance noise reduction and edge sharpness. In ultrasound imaging, the use of small patches helps reduce speckle noise while preserving essential details. By combining WNNM and homomorphic filtering, multiplicative noise is effectively converted into additive noise, making it easier to handle. This method is emphasized in studies such as \cite{sagheer2017ultrasound} (despeckling) and \cite{yang2021ultrasound} (tensor-based approach for 3D images). CT imaging, on the other hand, benefits from smaller patches to preserve fine details and edges that are crucial for accurate diagnosis in high-resolution images. Managing Poisson noise, especially at low doses, is a major concern. Techniques such as t-SVD and sparse coding with low-rank constraints, as demonstrated in \cite{sagheer2019denoising} and \cite{lei2018denoising}, help to balance noise reduction and edge sharpness. In ultrasound imaging, the use of small patches helps reduce speckle noise while preserving essential details. By combining WNNM and homomorphic filtering, multiplicative noise is effectively converted into additive noise, making it easier to handle. This method is emphasized in studies such as \cite{sagheer2017ultrasound} (despeckling) and \cite{yang2021ultrasound} (tensor-based approach for 3D images). For PET imaging, medium-sized patches balance capturing spatial and temporal detail. Non-local tensor low-rank methods, such as those in \cite{xie20193d}, utilize spatial and temporal correlations, improve image clarity, and reduce noise. In X-ray imaging, especially in dynamic sequences such as fluoroscopy, medium-size patches effectively maintain temporal consistency and manage quantum noise. Methods such as weighted local low-rank (\cite{hariharan2019preliminary}) and tensor SVD-based denoising (\cite{cheng2014curvilinear}) improve image quality while preserving anatomical details, making them well-suited for these applications.
\subsection{RQN4: Data Type}
The ability to handle and analyze different data types in healthcare is crucial due to the variety of information generated and collected in the medical field. Each data type offers unique insights, but deriving meaningful insights from these data types requires advanced technologies and methodologies. Advances in NLP, ML, and artificial intelligence (AI) have led to the development of techniques that can process, extract, and make sense of information from all three types of medical data (i.e., structured, semi-structured and unstructured data) to enable more comprehensive and informed healthcare decision making. For example, NLP can be used to extract diagnoses, treatments, and patient outcomes from physician notes, enabling better data integration and analysis \cite{demner2009can, velupillai2018using}. In contrast, ML techniques can be used to identify patterns, correlations, and anomalies in structured and semi-structured data \cite{habeeb2019real, sagiroglu2013big}. For example, ML algorithms can learn from historical patient data to predict disease advancement, suggest treatment options, or detect potential damaging events based on structured data. On the other hand, AI models can be used to process and analyze medical data across data types \cite{ma2020machine, johnson2021precision}. For example, image recognition using DL algorithms can help analyze medical images to detect signs of disease. AI-driven decisions can help healthcare professionals make more informed treatment decisions by analyzing a combination of structured and unstructured data. In Sections ~\ref{structured}, ~\ref{semistructured}, and ~\ref{unstructured} we discuss various data types not considered in the reviewed papers (Fig.~\ref{fg:SWork} shows unstructured data type that was considered in the reviewed publications). We also discuss whether it is possible to apply the LLORMA approximation technique to other data types, such as structured and semi-structured.
\subsubsection{Structured Data }{\label{structured}}
Structured data in healthcare refers to information that is organized and formatted in a consistent and predefined way so that it can be easily managed, accessed, and analyzed \cite{wang2018big}. Examples of structured data that were not included in
the publications reviewed include electronic health records (i.e., contain structured data such as patient demographics, medical history, medications, allergies, laboratory results, and clinical notes) and medical test results can be recorded in the form of a numeric or boolean value \cite{tayefi2021challenges}.

The LORMA matrix approximation technique can be applied to patient demographic data to analyze and extract insights from complex healthcare data. For example, if we have a comprehensive patient demographic dataset with various features (i.e., age, gender, ethnicity, medical history, and more), the study's goal is to reduce the dimensionality of the patient demographic dataset to visualize and cluster it. By representing the data as a matrix and applying matrix factorization, such as SVD, dimensionality can be effectively reduced while retaining the most important information. This facilitates the visualization of patient profiles in lower dimensions and makes it easier to identify patient clusters with similar demographic characteristics. These clusters can help healthcare professionals to understand similar patient characteristics better and adapt their services accordingly. The feasibility of this study depends on the availability of a sufficiently large and high-quality patient demographic dataset, as well as the privacy and ethical considerations associated with handling patient data. 
\subsubsection{Semi-structured Data}
\begin{figure}
    \centering
    \includegraphics[scale=0.45]{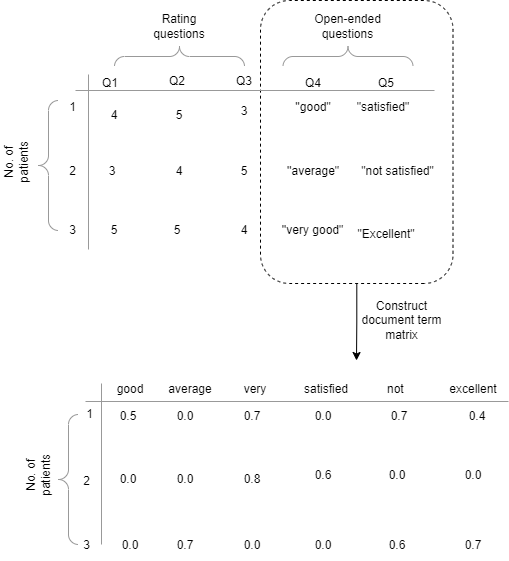}
    \caption{A demonstration of how a numeric matrix is constructed from a questionnaire with open-ended questions. In the DTM, each column represents a unique term or word that appears in the open-ended responses. These terms were processed and converted into a numerical representation using the frequency-inverse document frequency (TF-IDF). The numerical values in the DTM represent the TF-IDF weights for each term in each patient's response. TF-IDF evaluates how important a word is for a document (in this case, a patient's response) compared to a collection of documents (all patient responses). A higher TF-IDF value indicates a higher importance of the term in that specific response. For example, if patient 1 has higher TF-IDF values for terms such as \textit{Good}, \textit{Satisfied} and \textit{Great}, this means that these terms are more important in describing patient 1's response.}
    \label{fig:DTM}
\end{figure}

{\label{semistructured}}
Semi-structured data in healthcare refers to information that falls between fully structured and fully unstructured data \cite{zimmerman2008diagnostic}. They exhibit some form of organization but allow for more flexibility and variability compared to strictly structured data. Semi-structured data retains certain structural elements, such as predefined data fields or tags, while being able to accommodate different content and formats. Examples of semi-structured data that were not included in the publications reviewed include clinical forms (i.e., healthcare facilities use electronic forms for data collection that may allow free-text entries such as symptoms and medical history), radiology and laboratory reports (i.e., comments made by healthcare professionals), patient history summaries (i.e., may include information about the patient's current condition and treatment plan), medical literature, transcribed voice notes (i.e., healthcare providers may dictate notes during patient visits), and patient questionnaires (i.e., may include open-ended questions to collect patient feedback and information). The LORMA technique can be applied to a semi-structured dataset, for example, if a medical institution issues a patient satisfaction questionnaire that contains both open-ended questions and rating scales. The questionnaire may include questions on the quality of medical care, communication with healthcare providers and general satisfaction. Patients provide ratings on a scale of 1 to 5 for certain aspects and provide open-ended responses to capture additional feedback. To convert this questionnaire data into a matrix, a table in which each row corresponds to a patient and each column represents a specific question or rating category can be created, as shown in Fig.~\ref{fig:DTM}. The open-ended responses can be pre-processed by creating a document term matrix (DTM). A LORMA technique (e.g., PCA) can then be applied to the resulting matrix for feature reduction, especially if there are a large number of questions, making subsequent analyses more manageable.
 \subsubsection{Unstructured Data }{\label{unstructured}}
Unstructured healthcare data refers to information that does not follow a predefined and organized format, making it more difficult to manage, analyze, and interpret compared to structured data \cite{wang2018big}. Examples of unstructured data that were not included in the publications reviewed include clinical notes (i.e., details of a patient's medical history, symptoms, physical examinations and treatment plans), pathology reports (i.e., information about tissue samples and biopsies), handwritten documents (i.e., patient charts, prescriptions and other medical documents that are handwritten), audio recordings (i.e., dictated notes, voice memos, speech therapy sessions, lung sounds, recordings of heartbeats and audio recordings of patient conversations or surgical procedures that may contain valuable medical information), social media content and comments (i.e., patients may share their health experiences and ask for advice on social media platforms and online forums), and faxed versions of structured data (i.e., structured prescription data, when faxed to a pharmacy, loses its format and becomes an unstructured data) \cite{tayefi2021challenges}. The LORMA method is primarily used for image data and may not be directly applicable to audio data since audio data as a one-dimensional signal may not exhibit the same low-rank properties \cite{kong2020panns}. The concept of LORMA assumes that the data can be represented by a low-dimensional subspace, which aligns well with the structure of images \cite{lee2013local}. However, clinical notes, pathology reports, handwritten documents, and comments faxed versions of structured data may not have the same low-rank properties that LORMA techniques can exploit \cite{rusu2013converting}. In the context of social media content, latent semantic analysis (LSA) can be used to obtain a low-rank matrix approximation to discover underlying semantic patterns in a large collection of posts. For example, a dataset may consist of thousands of tweets. Each tweet can be considered as a document, and the words in these tweets form the terms. By constructing DTM, where rows represent tweets (i.e., number of tweets) and columns represent terms (i.e., words), a high-dimensional matrix is created that encodes the frequency of each word in each tweet. LSA reduces the dimensionality of this matrix by performing SVD, which helps to capture the latent semantic relations between terms and documents.
\subsection{RQN5: Applicable Only on Regular Data-type}
 The LORMA technique is useful in data analysis and dimensionality reduction, especially for regular data types such as matrices and tensors. However, it may not be directly applicable to irregular data types such as triangular matrices or irregularly distributed data points. The main reason for this limitation is that low-rank methods assume that the data can be well approximated by a low-dimensional subspace. Irregular data often violates this assumption since it may lack consistent patterns due to their unpredictable nature. This makes it challenging to identify a small set of underlying factors that can effectively capture the variation in the data. In the context of regular data with missing entries, techniques such as inpainting are used to fill in the missing values before applying low-rank methods. Inpainting is a process of estimating missing values based on the available data and certain assumptions about the underlying structure of the data. While inpainting can be effective, it comes with computational complexity and possible inaccuracies in prediction. Inpainting quality is critical when low-rank methods are used for data with missing entries. If the missing entries are not predicted accurately, it can significantly affect the performance of the LORMA. Inaccurate inpainting can introduce errors in the estimated low-rank structure, resulting in a poor representation of the original data.
 \subsection{RQN6: Shallow Similarity Methods}{\label{shallow}}
 The reviewed papers utilized different similarity measures (as shown in Fig.~\ref{fg:SWork} and discussed in Section~\ref{sec_1}) such as distance metrics (i.e., Euclidean distance \cite{khaleel2018denoising, liu2019medical, xie20193d, yang2020pet, yang2021ultrasound, chen2021joint, chen2021novel, shen2021ct, sagheer2017ultrasound, sagheer2019despeckling, sagheer2019denoising} and weighted Euclidean \cite{liu2015medical}), shallow learning models (i.e., KNN \cite{fu20163d, zhong2016predict, lei2018denoising, yang2018predicting, lv2019denoising} and k-means \cite{he2022denoising}) and image processing techniques (i.e., blocking matching \cite{zhao2022joint, wang2020modified, zhang2020image, xia2017denoising, sagheer2017ultrasound, zhai2018weighted} and weighted patch matching \cite{hariharan2018photon}) during the patch-matching stage to measure similar patches. These models are simple but effective techniques commonly used in different of machine-learning tasks. The use of shallow models in the papers under review indicates a preference for simplicity, interpretability, and computationally efficient techniques. These models provide simple and understandable solutions to a variety of machine-learning problems. 

While the above models are valuable techniques for measuring patch similarity, they also have certain limitations. For instance, Euclidean and weighted Euclidean distances are sensitive to the scale and range of the input features. If the features have different scales, this can lead to biased distance calculations and inaccurate similarity measurements. Normalizing the features can help mitigate this problem but may not always be sufficient for very heterogeneous data. KNN suffers from the curse of dimensionality, where performance decreases as the number of dimensions or features increases. In high-dimensional spaces, nearest neighbors may not provide accurate similarity estimates, resulting in lower performance. In addition, it can be challenging to determine the appropriate value of $k$, as a small $k$ can lead to noisy decisions, while a large $k$ can over-smooth local patterns. K-means is sensitive to the choice of initial cluster centroids and may converge to suboptimal solutions. The method also reaches its limits for non-uniformly sized or non-spherical clusters, as it assumes the same variance and spherical shape for all clusters. For complex medical image data, these assumptions may not hold, resulting in suboptimal cluster assignments and inaccurate similarity measurements. Block matching lacks fine-grained detail because it divides an image into fixed-size blocks, potentially missing subtle texture or structure variations that may exist within those blocks. This can lead to a loss of sensitivity to local variations, resulting in a coarse representation of the image content. Another significant drawback of block matching is its sensitivity to block size and placement. The choice of block size is critical because smaller blocks capture finer details but increase computational complexity, while larger blocks may miss crucial local information. On the other hand, weighted patch matching is sensitive to small variations in lighting, noise, or texture, which can cause patches to be classified as dissimilar even though they represent similar objects or structures. Additionally, it lacks a deep semantic understanding of the content within patches, making it unable to distinguish between patches with similar textures but different underlying objects or concepts.

The applicability of these shallow similarity checks is dependent on the specific problem and dataset characteristics. In some cases, more complex or deep learning (DL) models may be required to capture intricate patterns or achieve higher performance. In areas such as self-driving cars, for example, the complexity of processing large amounts of sensor data from cameras, lidar, and radar makes shallow methods for similarity measuring insufficient \cite{yeong2021sensor, miglani2019deep}. DL models such as convolutional neural networks (CNNs) and recurrent neural networks (RNNs) are remarkable for learning complex spatial and temporal features from raw sensor data, making them essential for self-driving systems \cite{gupta2021deep}. Similarly,  natural language processing (NLP) is another area where DL has made a significant impact. Tasks such as document similarity, sentiment analysis, and machine translation often require the analysis of large and complex textual data \cite{nandwani2021review}. DL models, including transformers and BERT, have revolutionized NLP by enabling models to learn complicated patterns and relationships in text data \cite{koroteev2021bert, ekman2021learning}. In these domains, the complexity and dimensionality of the data, as well as the need to capture complex patterns, make shallow model similarity checks inadequate. DL models have been shown to be more adaptable and effective, as they can automatically learn relevant features and relationships from data \cite{romero2013data, alzubaidi2021review}. As a result, researchers and practitioners in these domains are increasingly relying on DL for similarity and pattern recognition tasks \cite{youssef2020comparative, otter2020survey}. The shallow similarity methods discussed in this section would fail in these fields due to complexity, and only DL methods can adapt. The medical field is growing with a huge amount of data being collected and, therefore, might reach the stage where these shallow methods won't work anymore; therefore, this requires urgent investigation for DL methods in similarity check in medical image analysis
\subsection{RQN7: Effect of Patch Size in Medical Images}
When an image is divided into patches, the amount of local information captured in each patch is determined by the size of the patch. Smaller patch sizes are more focused on capturing fine details and local variations within a given region of interest. The ML algorithm can look more closely at smaller structures or subtle anomalies in a medical image by using smaller patches. This allows the algorithm to analyze localized patterns, textures, edges, or other specific features that may be required for accurate analysis. Smaller image patches can be useful in medical imaging, for example, to detect small lesions or specific anatomical structures that require a high level of detail. By analyzing smaller patches, the algorithm can better capture complex patterns that may go unnoticed in larger patches. Larger patch sizes, on the other hand, cover a broader context and capture more surrounding information in addition to the target region. This broader context provides a higher-level perspective and allows the algorithm to detect spatial relationships or global patterns in the image. Larger patches can assist in the analysis of larger structures or in capturing the overall layout of anatomical regions. Larger patch sizes in medical images can be useful for tasks like organ segmentation or classification, where the spatial arrangement of structures or the overall context is critical. The algorithm can achieve a better understanding of global features and relationships among different regions by incorporating more contextual information. 

The patch size chosen is determined by the scale and nature of the features or anomalies that the algorithm must detect or analyze. It is critical for the algorithm's performance and ability to extract meaningful information from medical images to strike a balance between capturing fine local details and incorporating sufficient contextual information. In the studied works, the patch sizes were varied depending on the noise level, i.e., with increasing noise, the patch size was increased \cite{khaleel2018denoising, sagheer2017ultrasound, sagheer2019despeckling, sagheer2019denoising, wang2020modified}. In contrast, \cite{he2019segmenting} varied patch size depending on the dimensions. Some authors \cite{xie20193d, ren2017drusen, chen2021joint, liu2019medical} compared different patch sizes and selected the one that provided optimal results. While others \cite{liu2015medical, xia2017denoising, zhong2016predict, hariharan2019preliminary, lv2019denoising, yang2020pet, zhang2020image, sagheer2017ultrasound, chen2021novel, shen2021ct, yang2018predicting, zhao2022joint} did not compare different patch sizes.  

To mitigate the effects of patch size, random search (RS) can be used in future studies to find the optimal patch size. RS is a valuable approach for hyper-parameter tuning, model selection, and optimization tasks because it provides a simple and effective way to explore the parameter space and determine appropriate parameter configurations for a given problem. In the analysis of medical images, such as MRI, CT scans, or ultrasound, the choice of patch size can significantly affect the performance of various tasks, such as segmentation, denoising or classification. The patch size determines the spatial context of the algorithm and can influence local and global feature extraction. By applying RS, one can systematically evaluate different patch sizes by drawing a random sample from a predefined range. In this approach, a patch size is selected, the algorithm is trained and evaluated with that patch size, and then the process is repeated with a different patch size. Depending on the task, the evaluation can be based on predefined performance metrics such as accuracy. The feasibility of RS in selecting the optimal patch size depends on several considerations. The size of the search space and the granularity of the patch size range are critical factors. If the search space is small or the range is well-defined, the RS can effectively cover the space with a reasonable number of iterations. However, in cases where the search space is large or the range is not well constrained, RS may require many iterations to sufficiently explore the space, making it less feasible. In addition, the computational cost of evaluating the algorithm for each patch size can impact feasibility. A hybrid approach can be used to improve feasibility. RS can be combined with other optimization techniques, such as Bayesian optimization, to leverage their strengths and overcome the limitations of RS. This hybrid approach enables more targeted exploration and can potentially increase efficiency in finding the optimal patch size.
\section{Conclusion and Future work}{\label{sec:recommendations}}
All publications included in this study utilized shallow models to measure the similarity between patches. The limitations of these techniques are discussed in detail in Section~\ref{shallow}. For future work, deep learning (DL) methods, such as DeepLab, can be explored to enhance the effectiveness of the proposed approaches. DeepLab is a DL-based semantic segmentation model commonly used for image analysis tasks, primarily to assign class labels to each pixel in an image, thereby identifying different objects and regions \cite{chen2017deeplab}. While DeepLab is not explicitly designed for measuring patch similarity, it can be leveraged for this purpose by extracting meaningful features and capturing the underlying semantic information within patches.

The process begins by extracting patches from the input image and isolating regions of interest. A pre-trained DeepLab model, which has been trained to recognize various objects and regions, is then applied to these patches to generate pixel-level semantic segmentation masks. These masks assign class labels to individual pixels, effectively delineating different objects and structures within the patches. The generated segmentation masks can serve as a basis for similarity measurement by quantifying the overlap between patches using Intersection over Union (IoU), a commonly used similarity metric \cite{hou2021improved}. The degree of similarity between patches can be assessed by calculating their IoU values, with a predefined threshold used to classify patches as either similar or dissimilar. Patches exceeding the threshold are considered similar, while those below it are regarded as dissimilar.

DeepLab offers the advantage of capturing underlying patterns and structures within patches, providing valuable context for similarity measurement. Due to its training on diverse datasets, DeepLab can effectively recognize complex objects and regions, making it particularly useful for scenarios involving intricate patterns and heterogeneous data. However, the feasibility of employing DeepLab for patch similarity measurement depends on several factors. The computational complexity of DeepLab is a potential challenge for applications requiring real-time performance or efficient processing, as it is a deep neural network demanding significant computational resources. Furthermore, training DeepLab necessitates a large dataset with pixel-level annotations, which may pose limitations for tasks with limited or heterogeneous training data. Another critical factor is selecting an appropriate similarity threshold to balance the trade-off between false positives and false negatives. Additionally, the choice of patch size is crucial, as small patches may fail to capture large-scale similarities, while larger patches may overlook fine-grained details.

In conclusion, there has been a significant shift toward the use of LLORMA in medical research since 2015, underscoring the efficiency of this method in capturing complex structures within medical data. Although shallow similarity measurement methods are widely adopted, they face challenges in handling the increasing complexity of healthcare data. Thus, exploring advanced techniques such as DeepLab for enhanced similarity measurement is recommended. This study also highlights the importance of applying LORMA to various healthcare data types and suggests incorporating reinforcement strategies (RS) to optimize patch size selection for medical image analysis in future research.

\section*{Data Availability Statement}
The datasets used and/or analysed during the current study are available from the corresponding author upon reasonable request.
\section*{Conflict of Interest}
The authors declare no conflict of interest.
\bibliography{sn-bibliography}
\end{document}